\documentclass[journal]{IEEEtran} 

\usepackage{amsmath,amsfonts}
\usepackage{array}
\usepackage{textcomp}
\usepackage{stfloats}
\usepackage{url}
\usepackage{graphicx}
\usepackage{MnSymbol}
\usepackage{color}
\usepackage{comment}
\usepackage{acronym}
\usepackage{caption}
\usepackage{subcaption}
\usepackage{empheq}
\usepackage[font=small]{caption}
\usepackage{enumitem}
\usepackage{cite}

\newcommand{\bs}[1]{\boldsymbol{#1}}
\newcommand{\x}{\bs{x}}
\newcommand{\z}{\bs{z}}
\newcommand{\y}{\bs{y}}
\newcommand{\A}{\bs{A}}

\newcommand{\rhou}[1]{\varrho_{#1}}
\newcommand{\omegau}[1]{\tilde{\bs{\omega}}_{#1}}
\newcommand{\jj}{\tilde{\bs{\jmath}}}
\newcommand{\den}[2]{\zeta^2_{#1} + \zeta^2_{#2}}

\acrodef{pll}[PLL]{Phase-Locked Loop}
\acrodef{qss}[QSS]{Quasi Steady-State}
\acrodef{dfig}[DFIG]{Doubly-Fed Induction Generator}
\acrodef{dft}[DFT]{Discrete Fourier Transform}
\acrodef{pmu}[PMU]{Phasor Measurement Unit}
\acrodef{rl}[RL]{Reinforcement Learning}
\acrodef{lti}[LTI]{linear time-invariant}

\begin{document}
\bstctlcite{IEEEexample:BSTcontrol}

\title{Complex Frequency as Generalized Eigenvalue}

\author{Nikolaos Sofos,~\IEEEmembership{Student Member,~IEEE,}   and  Federico Milano,~\IEEEmembership{Fellow,~IEEE}
  \thanks{N.~Sofos and F.~Milano are with the School of Elec.~\& Electron.~Eng., University College Dublin, Dublin, D04V1W8, Ireland.  \\ E-mails: nikolaos.sofos1@ucdconnect.ie, federico.milano@ucd.ie}%
\thanks{This work is supported by the Environmental Protection Agency~(EPA) under project CRIE, Grant No.~2024-CE-1266.}
\vspace{-2mm}
}

\maketitle

\begin{abstract}
This paper shows that the concept of complex frequency, originally introduced to characterize the dynamics of signals with complex values, constitutes a  generalization of eigenvalues when applied to the states of \ac{lti} systems.  Starting from the definition of geometric frequency, which provides a geometrical interpretation of frequency in electric circuits that admits a natural decomposition into symmetric and antisymmetric components associated with amplitude variation and rotational motion, respectively, we show that complex frequency arises as its restriction to the two-dimensional Euclidean plane. For \ac{lti} systems, it is shown that the complex frequencies computed from the system's states subject to a non-isometric transformation, coincide with the original system's eigenvalues. This equivalence is demonstrated for diagonalizable systems of any order.   The paper provides a unified geometric interpretation of eigenvalues, bridging classical linear system theory with differential geometry of curves.  The paper also highlights that this equivalence does not generally hold for nonlinear systems.  On the other hand, the geometric frequency of the system can always be defined, providing a geometrical interpretation of the system flow.  A variety of examples based on linear and nonlinear circuits illustrate the proposed framework. 
\end{abstract}


\vspace{-2mm}

\section*{Notation}
\label{sec:notation}

\textit{General Notation:}
\vspace{-0.4em}
\begin{tabbing}
xxxxxxxx \= xxxxxxxxxxxxxxxxxxxxxxxxxxxxxxxxx \kill
\hspace*{0.96em}$a, A$ \> scalars \\
\hspace*{0.96em}$\bar{a}$ \> complex numbers \\
\hspace*{0.96em}$\bs{a}$ \> vectors \\
\hspace*{0.96em}$\bs{\tilde{a}}$ \> bivectors \\
\hspace*{0.96em}$\bar{\bs{a}}$ \> vectors with complex entries\\
\hspace*{0.96em}$\bs{A}$ \> matrices\\
\hspace*{0.96em}$\bar{\bs{A}}$ \> matrices with complex entries\\
\hspace*{0.96em}$\bs{a^\top}$ \> transpose of a vector \\
\hspace*{0.96em}$\bs{A^\top}$ \> transpose of a matrix \\
\hspace*{0.96em}$a'$ \> time derivative of a quantity $a$
\end{tabbing}

\textit{Scalars:}
\vspace{-0.4em}
\begin{tabbing}
xxxxxxxx \= xxxxxxxxxxxxxxxxxxxxxxxxxxxxxxxxx \kill
\hspace*{0.96em}$i$ \> current \\
\hspace*{0.96em}$t$ \> time \\
\hspace*{0.96em}$v$ \> voltage\\
\hspace*{0.96em}$\rho$ \> symmetric term of geometric frequency \\
\hspace*{0.96em}$|\tilde{\bs{\omega}}|$ \> norm of $\tilde{\bs{\omega}}$  \\
\end{tabbing}

\vspace{-2mm}

\textit{Complex numbers:}
\vspace{-0.4em}
\begin{tabbing}
xxxxxxxx \= xxxxxxxxxxxxxxxxxxxxxxxxxxxxxxxxx \kill
\hspace*{0.96em}$\bar\lambda_i$ \> $i$-th eigenvalue of a A\\
\hspace*{0.96em}$\bar\lambda_i^*$ \> complex conjugate of the eigenvalue $\bar\lambda_i$ 
\end{tabbing}

\textit{Vectors:}
\vspace{-0.4em}
\begin{tabbing}
xxxxxxxx \= xxxxxxxxxxxxxxxxxxxxxxxxxxxxxxxxx \kill
\hspace*{0.96em}$\bs{e}_i$ \> $i$-th vector of an orthonormal basis \\
\hspace*{0.96em}$\bs{l}_i$ \> left eigenvector of $\A$ corresponding to $\bar\lambda_i$ \\
\hspace*{0.96em}$\bs{r}_i$ \> right eigenvector of $\A$ corresponding to $\bar\lambda_i$ \\
\hspace*{0.96em}$\bs{u}$ \> generalized velocity \\
\hspace*{0.96em}$\bs{x}$ \> generalized position \\
\hspace*{0.96em}$\bs{\zeta}$ \> transformed states after using the isomorphism\\
\hspace*{0.96em}$\bar{\bs{\xi}}$ \> transformed states
\end{tabbing}

\textit{Matrices:}
\vspace{-0.4em}
\begin{tabbing}
xxxxxxxx \= xxxxxxxxxxxxxxxxxxxxxxxxxxxxxxxxx \kill
\hspace*{0.96em}$\bs{A}$ \> state matrix \\
\hspace*{0.96em}$\bs{D}$ \> symmetric part of $\bs{G}$ \\
\hspace*{0.96em}$\bs{G}$ \> matrix similar to $\A$ \\
\hspace*{0.96em}$\bs{Q}$ \> antisymmetric part of $\bs{G}$ \\
\hspace*{0.96em}$\bar{\bs{U}}$ \> matrix whose columns are the eigenvectors of $\bs{A}$ \\
\hspace*{0.96em}$\bs{W}$ \> transformation matrix \\
\hspace*{0.96em}$\bar{\bs{\Lambda}}$ \> diagonal matrix with entries the eigenvalues of $\A$ 
\end{tabbing}

\textit{Bivectors:}
\vspace{-0.4em}
\begin{tabbing}
xxxxxxxx \= xxxxxxxxxxxxxxxxxxxxxxxxxxxxxxxxx \kill
\hspace*{0.96em}$\jj$ \> imaginary unit  \\
\hspace*{0.96em}$\tilde{\bs{\omega}}$ \> antisymmetric component of geometric frequency 
\end{tabbing}

\section{Introduction}
\label{sec:intro}

\subsection{Motivation}

It is well known that the stability of the equilibria of systems of linear differential equations is defined by the eigenvalues of the state matrix of the system.  On the other hand, complex frequency can provide information about the dynamics of complex quantities, such as the voltage of a bus in a power system \cite{milano2021complex}.   The research question that is addressed in this work is whether there is a link between complex frequency and eigenvalues and, if there is such a link, what are the conditions under which there is a perfect equivalence.  We answer both questions and show that eigenvalues are complex frequencies of a non-isomorphic coordinate transformation of the system state space.

\vspace{-2mm}
\subsection{Literature Review}

The analysis of the dynamical behavior of electric circuits has been an active area of research for several decades.  Early works in nonlinear circuit theory demonstrated that even relatively simple electronic circuits can exhibit a wide range of dynamical phenomena, including oscillations, bifurcations and chaotic behavior.  Nonlinear circuit theory is extensively discussed in \cite{chua1987linear}, where electrical networks are modeled as systems of nonlinear differential equations.  In particular, \cite{kennedy2002threeI} and \cite{kennedy2002threeII} provide a systematic framework for understanding the transition from linear dynamics to nonlinear oscillatory and chaotic regimes in electronic circuits. Similarly, \cite{kennedy2002chaos} and \cite{murali2002simplest} demonstrate that even low-order and classical oscillator circuits can exhibit chaotic behavior under specific conditions.  Several works have also focused on the analysis of the stability of nonlinear circuits.  For instance, \cite{yang2012lyapunov} studies Lyapunov stability and passivity conditions for nonlinear circuits, while \cite{anghel2013algorithmic} proposes algorithmic methods for constructing Lyapunov functions for power system stability analysis.

In the context of the study of the stability of the flows of a dynamic system, \cite{ginoux2009differential} studies the application of differential geometry to dynamical systems.  It is shown that the stability of fixed points of systems of order up to three can be analyzed through their flow curvature manifold.  More recently, this approach has been applied to \ac{lti} systems in \cite{wang2020torsion}, where the eigenvalues of a system's Jacobian matrix are related to the geometric properties of its trajectory.  In particular, asymptotic stability can be assessed by evaluating the torsion and higher-order curvatures of the phase-space trajectory.

Most studies on \ac{lti} systems have traditionally relied on algebraic methods, specifically the determination of their eigenvalues and eigenvectors \cite{chen1984linear, hespanha2018linear}.  However, attempts have also been made to establish a geometrical interpretation of the behavior of these systems, which led to a deeper understanding of their dynamics.
Foundations for this perspective are laid in \cite{arnold2012geometrical}, where it is established that ordinary differential equations can be interpreted geometrically by treating the states of the system as points on a manifold and their time derivatives as continuous vectors.  This viewpoint is further developed in \cite{guckenheimer2013nonlinear}, where the movement of eigenvalues in the complex plane is connected to the trajectories in state space.

Efforts to extract useful information from the complex-plane representation of the state-space flows of linear systems can be found in \cite{macfarlane1968representation}.  This reference shows that the trajectory of a linear constant-coefficient second-order system can be mapped onto the complex plane, where the flow is represented by a circle.  This geometric transformation, which is not limited to second-order systems, allows the dynamic behavior of the system, captured within the structure of the system matrix, to be visualized as a specific locus.
These loci provide information about the stability and oscillatory behavior that the system may exhibit.  It has also been noted that this mapping is useful in the design of multivariable feedback controllers.  Building upon these geometrical interpretations, \cite{macfarlane1969use} further analyzes the connection between the eigenvalues of a linear system and components of its power.  

Recent works have focused on the development of concepts that provide additional insight into the dynamical behavior of electrical circuits.  In particular, reference \cite{milano2021complex} introduces a formulation, called \textit{complex frequency}, that allow extracting additional information about frequency at power system buses.  The imaginary part of the complex frequency corresponds to variations of the local bus frequency, while the real part is associated with changes in voltage magnitude.

While the complex frequency concept has provided several insights to the study of ac circuits, in particular, power systems, the fact that complex frequency can be utilized only for two-dimensional vector fields, constitutes a significant limitations.  To move from two to higher dimensional spaces requires the adoption of Clifford algebra, also known as geometric algebra. 

This algebra has found relevant applications in physics and engineering.  For example, \cite{Jancewicz:1989} presents a comprehensive rigorous treatment of electrodynamics based on Clifford algebra.  
Reference \cite{castro2010use} utilizes Clifford algebra in circuit analysis and proposes a geometric interpretation of electrical power quantities.  More recently, \cite{eid2022geometric} proposed a procedure for obtaining a multivector representation of the geometric angular frequency in multiphase electrical systems, known as the Darboux bivector.  Geometric approaches have also been extensively developed in signal processing and image analysis.  In particular, \cite{felsberg2001monogenic} introduces the concept of the monogenic signal as a 2D extension of the analytic signal.  Building upon this framework, \cite{wietzke2008differential} studies the differential geometry of monogenic signal representations, establishing links between geometric structures and local signal features in two-dimensional data.  More recently, \cite{Dalal:2025} has shown that the eigenfuntions of rotational invariant LTI system can be described with bivectors, which is the natural way to define a rotation in geometric algebra.

Leveraging the tools provided by Clifford algebra, reference \cite{milano2021geometrical} generalizes the concept of complex frequency and introduces that concept of \textit{geometric frequency} as a multivector composed of two terms: a symmetric component and an antisymmetric component.  This work also establishes a connection between the geometric frequency and the instantaneous power of a circuit.
In the same vein, reference \cite{milano2021frenet} provides a geometrical interpretation of electrical quantities and derives several expressions that link the time derivatives of voltage, current, and frequency in electrical circuits with the Frenet frame. A related perspective is presented in \cite{milano2024instantaneous}, where instantaneous power theory is revisited using concepts from classical mechanics, further highlighting the connection between electrical circuit dynamics and geometric interpretations of energy and power.  Recently, \cite{garcia2026instantaneous} analyzes the conceptual differences between geometric frequency, complex frequency, and instantaneous frequency formulations, providing a clarification of the utility and proper use of each concept.

\subsection{Contributions}

The main contribution of this work is to establish the fundamental relationship between complex and geometric frequency, and the eigenvalues of \ac{lti} systems.  
Specifically, this paper provides a formal proof that the complex frequency computed from a system's states, when subjected to a specific transformation, is mathematically equivalent to the system's eigenvalues.
This equivalence indicates that, for \ac{lti} systems, the information about the stability of a dynamic system is embedded in the geometrical definition of complex frequency.  

The paper also shows that the equivalence between eigenvalues and complex frequency does not hold for nonlinear systems.  However, since geometric frequency is also a property of the system's state-space trajectory, it remains well-defined and interpretable for nonlinear systems, whereas eigenvalues generally lack meaning during transients.

The second key contribution of the paper is to assume that the states of a system of differential equations are \textit{generalized positions}, say $\boldsymbol x$.  In this way, regardless the fact that the differential equations that describe the dynamics of the circuit are linear or not, the link between generalized accelerations $\boldsymbol{x}''$ and generalized velocities $\boldsymbol{x}'$ is always linear, although possibly time-varying. 

\subsection{Paper Organization}

This paper is organized as follows. Section \ref{sec:outlines} recalls relevant concepts of differential geometry and the definition of complex and geometric frequencies, and introduces the connection between complex numbers and a specific class of matrices.
Section \ref{sec:connection} explains how the quantities defined in differential geometry can be applied to systems of Ordinary Differential Equations (ODEs).  In particular,  Section \ref{sec:connection} considers first the connection between the eigenvalues of \ac{lti} and nonlinear systems and the complex frequency of their state variables is introduced and then illustrates the properties of the geometric frequency for nonlinear ODEs.  Section \ref{sec:examples} presents a variety of analytical examples that prove the observed relationship between complex frequency and eigenvalues both in linear and nonlinear systems.  Conclusions and future work are given in Section \ref{sec:conclusions}.


\section{Background}
\label{sec:outlines}

This section recalls the definitions of relevant quantities that are utilized in developments of the paper. These key concepts include geometric frequency,  complex frequency and the isomorphism between complex numbers and a specific class of matrices.  It is also shown that the complex frequency is a specific case of geometric frequency in two dimensions.

\subsection{Geometric Frequency}

Reference \cite{milano2021geometrical} proposes a geometrical interpretation of frequency in electric circuits.  According to this interpretation, the frequency is defined as a multivector with symmetric and antisymmetric components. More specifically, the geometric frequency of a generalized velocity $\bs{u} \in \mathbb{R}^n$ can be decomposed into a symmetric radial term:
\begin{equation}
    \label{eq:geom_freq_sym}
    \rhou{u} =  \frac{\bs{u} \cdot \bs{u}'}{|\bs{u}|^2}  \,, 
\end{equation}
and an antisymmetric rotating term:
\begin{equation}
    \label{eq:geom_freq_antisym}
    \omegau{u} =  \frac{\bs{u} \wedge \bs{u}'}{|\bs{u}|^2}  \,, 
\end{equation}
so that the full geometric frequency can be written as:
\begin{equation}
\label{eq:geom_freq}
\rhou{u} + \omegau{u}
= \frac{\bs{u}\bs{u}'}{|\bs{u}|^2} \, .
\end{equation}
Finally, the generalized velocity is defined as the time derivative of the generalized position $\x$, namely
\begin{equation}
\label{eq:position_to_velocity}
\bs{u} = \x' \, .
\end{equation}
The definitions of the Euclidean norm $|\bs{u}|$ and of the products $\bs{u} \bs{u}'$, $\bs{u} \cdot \bs{u}'$ and $\bs{u} \wedge \bs{u}'$ are given in the Appendix A.
As the remainder of this work focuses on the analysis of the geometric frequency of the generalized velocity vector field $\bs{u}$, we assume that the original states $\x$ are smooth and differentiable at least twice.

It is important to note that, while the geometric frequency can be defined for any time-dependent vector, if it is calculated based on the velocity vector field, then $\varrho$ and $|\tilde{\bs{\omega}}|$ are geometrical \textit{geometrical invariants}, that is, their value is the same independently from rotation or translation (e.g., isometric transformations) of the coordinate system that is utilized to define the components of the velocity vector field \cite{stoker1969differential}.  
On the other hand, geometric invariants change for transformations that do not preserve distances, such as scaling and affine transformations.  This is a key point of the paper, as we show below that the transformation that leads to obtain the eigenvalues of a linear system is not isometric.

\subsection{One- and Two-Dimensional Cases}

Geometric frequency, as defined in \cite{milano2021geometrical}, can be calculated for any generalized velocity $\bs{u} \in \mathbb{R}^n$. 

If $\bs{u} \in \mathbb{R}$, then let $\bs{u} = [u_1]$.
Then:
\begin{equation}
    \label{eq:geom_freq_one_by_one}
    \rhou{u} + \omegau{u} = \frac{u_1 u_1'}{u_1^2} + 0 = \frac{u_1'}{u_1} \,.
\end{equation}
Hence, if $\bs{u} \in \mathbb{R}$, its geometric frequency consists only of the scalar component.

If $\bs{u} \in \mathbb{R}^2$, then let $\bs{u} = [u_1, u_2]$.
Then:
\begin{equation}
    \label{eq:geom_2D}
    \rhou{u} + \omegau{u} = \frac{u_1 u_1' + u_2 u_2'}{u_1^2 + u_2^2} + \frac{u_1u_2' - u_2u_1'}{u_1^2 + u_2^2}(\bs{e}_1 \wedge \bs{e}_2) \,.
\end{equation}
In \cite{milano2021geometrical} is shown that, in fact, the algebra of complex numbers is equivalent to the two-dimensional Clifford algebra ${\rm{Cl}_{0,1}}(\mathbb{R})$ \cite{Jancewicz:1989}.  Thus, the unit bivector $\bs{e}_1 \wedge \bs{e_2}$  and  the classical imaginary unit $\jj$ are algebraically isomorphic in the 2D Euclidean plane. For this reason, these two are treated as operationally interchangeable in the paper.  We can thus rewrite \eqref{eq:geom_2D} as:
\begin{equation}
    \label{eq:geom_2D_complex}
\begin{aligned}
    \rhou{u} + \omegau{u} &= \frac{u_1 u_1' + u_2 u_2'}{u_1^2 + u_2^2} + \frac{u_1u_2' - u_2u_1'}{u_1^2 + u_2^2}\jj \\ 
    &= \frac{|\bs{u}|'}{|\bs{u}|} + \theta'\jj \, ,
\end{aligned}
\end{equation}
where $\theta =  \arctan(u_2/u_1)$.
Hence, the geometric frequency of a two-dimensional vector is equivalent to a complex quantity.  In the following, to help distinguish between two-dimensional systems and multi-dimensional cases with $n>2$, when considering two dimensional vectors, we refer to their geometric frequency as \textit{complex frequency}.

\subsection{Matrix Representation of Complex Numbers}

Part of our analysis uses the well-known isomorphism between the field of complex numbers $(\mathbb{C})$ and a specific class of real matrices $(\bs{M}_2(\mathbb{R})
)$, as follows:
\begin{equation}
    \label{eq:isomorphism}
    \phi : \mathbb{C} \rightarrow \bs{M}_2(\mathbb{R}),  \qquad \phi(a + b \jj) = \begin{bmatrix}
        a & -b  \\ 
        b & a
    \end{bmatrix} \,.
\end{equation}

This isomorphism preserves algebraic operations, and it also allows certain matrix quantities to be interpreted analogously to familiar complex-number notions \cite{needham2023visual, strang2022introduction, nielsen2010quantum}.  In particular, it is straightforward to observe that $|\bar{z}|^2 = {\rm det}(\phi(\bar{z}))$.

\section{Eigenvalues as Complex Frequencies}
\label{sec:connection}

In the first part of this section, we define the conditions under which eigenvalues can be interpreted as complex frequencies of generalized velocities.  In the second part, we show how the equivalence between eigenvalues and complex frequency does not hold, in general, for nonlinear systems and we illustrate the behavior of geometric frequency of the system for trajectories approaching an equilibrium.

\subsection{Linear systems}
Consider the case of a system of \ac{lti} homogeneous ODEs:
\begin{equation}
    \label{eq:linear_system}
    \x'=\bs{A} \x ,
\end{equation}
where $\x \in \mathbb{R}^n$ and $\A \in \mathbb{R}^{n\times n}$ is 
time-invariant matrix. \\ 
Then, the time derivative of \eqref{eq:linear_system} gives:
\begin{equation}
    \label{eq:sec_ord_linear_system}
    \x''=\bs{A} \x' \, ,
\end{equation}
or, equivalently:
\begin{equation}
    \label{eq:system_speed}
    \bs{u}' = \A \bs{u} \, .
\end{equation}

If $\A$ is diagonalizable, one can consider the decomposition:
\begin{equation}
    \label{eq:decomposition}
    \A = \bar{\bs{U}}^{-1}\bar{\bs{\Lambda}}\bar{\bs{U}} \,,
\end{equation}
where $\bar{\bs{U}}$ is a matrix whose columns are the eigenvectors of $\bar{\bs{\Lambda}}$ and $\bar{\bs{\Lambda}}$ is a diagonal matrix whose diagonal entries are all the eigenvalues of $\A$, regardless of whether they are real or complex, as follows:
\begin{equation}
    \label{eq:Lambda_diag}
\begin{aligned}
    \bar{\bs{\Lambda}} = \mathrm{diag}\big(
    \bar{\lambda}_1, \ldots, \bar{\lambda}_r,\,
    \bar{\lambda}_{r+1}, \ldots, \bar{\lambda}_{r+2s}
    \big)
    \,,
\end{aligned}
\end{equation}
where:
\begin{itemize}
    \item $r$ real eigenvalues 
    \begin{equation*}
    \begin{aligned}
        \bar{\lambda}_{i} = \mu_i \, , \quad  i = 1, \dots, r \, , 
    \end{aligned}
    \end{equation*}
    \item $s$ pairs of complex conjugate eigenvalues
    \begin{equation*}
    \begin{aligned}
        \bar\lambda_{r+2k-1}& = \alpha_{k} + \beta_{k} \jj \, ,  \\ \bar\lambda_{r+2k} &= \alpha_{k} - \beta_k \jj \, , \quad \beta_{k} \neq 0 \, , \quad k = 1, \dots, s \, , 
    \end{aligned}
    \end{equation*}
\end{itemize}
with $r + 2s =n$.  As the system matrix $\A$ is real-valued, the characteristic polynomial has real coefficients. Hence, its non-real roots occur in complex conjugate pairs \cite{horn2012matrix}. The ordering $\bar\lambda_{r+2k}$, $\bar\lambda^{*}_{r+2k-1}$ is a conventional sorting.

Then, let:
\begin{equation}
    \label{eq:xi}
    \bar{\bs{\xi}} = \bar{\bs{U}} \bs{u} \,,
\end{equation}
where $\bar{\bs{\xi}}$ is complex because the eigenvectors that form matrix $\bar{\bs{U}}$ can in general be complex. 

Then, the dynamic system can be written as:
\begin{equation}
    \label{eq:xi_dynamic}
    \bar{\bs{\xi}}' = \bar{\bs{\Lambda}} \bar{\bs{\xi}} \,, 
\end{equation}
The detailed derivation of this result is provided in Appendix~\ref{app:xi_dynamics}.

This system has $r$ real and $s$ complex variables, which define $r+4s$ real quantities, since each pair of complex eigenvalues is characterized by four real quantities.  However, since the original system is an $r+2s$ order system and real, only $2s$ of the $4s$ quantities are effectively independent. In fact, for each pair of complex conjugate eigenvalues, $\bar\lambda_{r+2k} = \bar\lambda_{r+2k-1}^*$.

Then, by utilizing the isomorphism between complex numbers and $2 \times 2$ matrices that was introduced in Section \ref{sec:outlines}, we can rewrite \eqref{eq:xi_dynamic} as an $r+2s$ order real system in the form:
\begin{equation}
    \label{eq:zeta_dynamic}
    \bs{\zeta}' = \bs{G\zeta}
    =
    \left[
    \begin{array}{ccc|cccc}
    \mu_1 &        &       &            &            &          &          \\ 
          & \ddots &       &            &            &          &          \\ 
          &        & \mu_r &            &            &          &          \\ \hline
          &        &       & \alpha_{1} & -\beta_{1} &          &          \\ 
          &        &       & \beta_{1}  & \alpha_{1} &          &          \\ 
          &        &       &            &            & \ddots   &          \\ 
          &        &       &            &            & \alpha_s & -\beta_s \\ 
          &        &       &            &            & \beta_s  & \alpha_s
    \end{array}
    \right]
    \bs{\zeta} \,,   
\end{equation}
where for each real eigenvalue, the corresponding state is one–dimensional and given by:
\begin{equation}
    \boldsymbol{\zeta}_{i} = \zeta_i\,\mathbf e_1 \,,
\end{equation}
and for each pair of complex conjugate eigenvalues \mbox{$\alpha_k\pm\beta_k\jj$}, 
$k=1,\dots,s$, the corresponding real realization is two–dimensional and given by:
\begin{equation}
    \boldsymbol{\zeta}_{c,k} =
    \zeta_{r+2k-1}\,\mathbf e_1
    +
    \zeta_{r+2k}\,\mathbf e_2 \,.
\end{equation}
The complex transformed state vector $\bar{\boldsymbol{\xi}}$ is defined consistently with the modal decomposition.
For each real eigenvalue $\bar\lambda_i$, $i=1,\dots,r$, the corresponding state variable is real and given by:
\begin{equation}
    \bar{\bs{\xi}}_i = \zeta_i \,.
\end{equation}
For each pair of complex conjugate eigenvalues $\alpha_k \pm \beta_k\jj$, \mbox{$k=1,\dots,s$}, the corresponding complex state variable is defined as:
\begin{equation}
    \bar{\bs{\xi}}_{r+2k-1}
    =
    \zeta_{r+2k-1}
    +
    \zeta_{r+2k}\jj,
    \qquad
    \bar{\bs{\xi}}_{r+2k}
    =
    \bar{\bs{\xi}}_{r+2k-1}^{*} \,.
\end{equation}
Then, we can write:
\begin{equation}
    \label{eq:decomposition_matrix_real}
    \A = \bs{W}^{-1}\bs{G}\bs{W} \,,
\end{equation}
with:
\begin{equation}
    \label{eq:transformation}
    \bs{\zeta} = \bs{W} \bs{u} \,.
\end{equation}

Consider again the system given by \eqref{eq:system_speed}, with eigenvalues given by \eqref{eq:Lambda_diag}.  Our goal is to show that there exists a real, invertible matrix $\bs{W}$ such that \eqref{eq:decomposition_matrix_real}, where $\bs{G}$ is block diagonal with $1 \times 1$ and $2 \times 2$ real blocks corresponding to real and complex eigenvalues, respectively.

Let $\bar\lambda_i \in \mathbb{R}$ be a real eigenvalue of $\bs{A}$ with associated real left eigenvector:
\begin{equation}
    \label{eq:real_eigenvectors}
    \bs{w}_{ri}\bs{A} = \bar\lambda_i\bs{w}_{ri} \,, \qquad \bs{w}_{ri} \in \mathbb{R}^n \,.
\end{equation}
The corresponding block to this eigenpair in $\bs{G}$ is:
\begin{equation}
    \bs{G}_i = [\mu_i] \,.
\end{equation}

Then, for each complex eigenvalue pair indexed by $k$, let:
\begin{equation}
    \bs{w}_{ck}\bs{A}  = (\alpha_k + \beta_k \jj)\bs{w}_{ck} \,, \qquad \bs{w}_{ck} \in \mathbb{C}^n \,.
\end{equation}
We can then write the eigenvector as:
\begin{equation}
    \bs{w}_{ck} = \z_{ck} + \y_{ck} \jj\,, \qquad \z_{ck} , \y_{ck} \in \mathbb{R}^n \,.
\end{equation}
Substituting into the eigenvalue equation and equating real and imaginary parts yields:
\begin{equation}
    \label{eq:complex_real_imag_k}
    \begin{aligned}
    \z_{ck}\A  &= \alpha_k \z_{ck} - \beta_k \y_{ck}, \\ 
    \y_{ck}\A  &= \beta_k \z_{ck} + \alpha_k \y_{ck},
    \end{aligned}
    \qquad k = 1, \dots, s .
\end{equation}
The corresponding block to this eigenpair in $\bs{G}_{r+k}$ is:
\begin{equation}
    \bs{G}_{r+k} = \begin{bmatrix}
        \alpha_k & -\beta_k \\ 
        \beta_k & \alpha_k
    \end{bmatrix} \,.
\end{equation}
Now, we define the real matrix:
\begin{equation}
    \bs{W} =
    \begin{bmatrix}
    \bs{w}_{ri} & \cdots & \bs{w}_{rr} &
    \z_{c1} & \y_{c1} &
    \cdots &
    \z_{cs} & \y_{cs}
    \end{bmatrix} ^{\bs{\top}}
     \,,
\end{equation}
where $\bs{W} \in \mathbb{R}^{n\times n}$. 

Finally, we define:
\begin{equation}
    \bs{G} =
    \mathrm{diag}
    \left(
    \mu_1, \dots, \mu_r,
    \begin{bmatrix}
        \alpha_1 & -\beta_1 \\ 
        \beta_1 & \alpha_1
    \end{bmatrix},
    \dots,
    \begin{bmatrix}
        \alpha_s & -\beta_s \\ 
        \beta_s & \alpha_s
    \end{bmatrix}
    \right) \,.
\end{equation}

Using both \eqref{eq:real_eigenvectors} and \eqref{eq:complex_real_imag_k}, it follows that:
\begin{equation}
    \bs{W A} = \bs{G W} \,.
\end{equation}
As $\bs{W}$ is invertible, \eqref{eq:decomposition_matrix_real} is verified.

We have thus shown that matrix $\A$ can be decomposed into decoupled real dynamics via a real similarity transformation. Each real eigenvalue of $\A$ generates an independent first-order subsystem, while each complex conjugate pair generates an independent second-order real subsystem.

For each independent first-order subsystem $\bs{\zeta}_i' = \mu_i\zeta_i$, we have that:
\begin{equation}
    \mu_i = \frac{\zeta_i'}{\zeta_i} \,,
\end{equation}
which shows that the eigenvalue of the system coincides with the definition of the real part of the complex frequency, as shown in \eqref{eq:geom_freq_one_by_one}.

For each independent second-order subsystem we have that:
\begin{equation}
    \bs\zeta_{c,k}' = 
    \begin{bmatrix}
            \zeta_{r+2k-1}' \\  \zeta_{r+2k}'
    \end{bmatrix} = 
    \begin{bmatrix}
            \alpha_k  & -\beta_k \\ \beta_k & \alpha_k
        \end{bmatrix}\begin{bmatrix}
            \zeta_{r+2k-1} \\ \zeta_{r+2k}
    \end{bmatrix} \, ,
\end{equation}
which can be decomposed into a symmetric and anti-symmetric part:
\begin{equation}
    \label{eq:zeta_dynamic_decomposed}
    \bs\zeta_{c,k}' =  \bs{D}\bs\zeta_{c,k} +  \bs{Q}\bs\zeta_{c,k} \,,
\end{equation}
where
\begin{equation}
  \bs{D} = \frac{1}{2}(\bs{G}_{r+k} + \bs{G^\top}_{r+k})  \,, \quad \bs{Q} = \frac{1}{2}(\bs{G}_{r+k} - \bs{G^\top}_{r+k})  \,,
\end{equation}
thus:
\begin{equation}
    \label{eq:zeta_decomp_matrices}
    \bs{D}  =  \begin{bmatrix}
        \alpha_k  & 0  \\ 
        0 & \alpha_k
    \end{bmatrix} \,,  \quad \bs{Q} =  \begin{bmatrix}
        0  & -\beta_k  \\ 
        \beta_k & 0
    \end{bmatrix} \,.
\end{equation}

Then, by applying the definition of complex frequency directly to the dynamic system as it was defined in \eqref{eq:zeta_dynamic} and decomposed in \eqref{eq:zeta_dynamic_decomposed}:
\begin{equation}
    \label{eq:zeta_complex_real}
    \frac{\bs\zeta_{c,k} \cdot \bs\zeta_{c,k}'}{|\bs\zeta_{c,k}|^2} = \frac{\bs\zeta_{c,k}^{\bs{ \top}} \bs{D} \bs\zeta_{c,k}}{|\bs\zeta_{c,k}|^2} = \alpha_k,
\end{equation}
and 
\begin{equation}
    \label{eq:zeta_complex_imag}
    \frac{\bs\zeta_{c,k} \wedge \bs\zeta_{c,k}'}{|\bs\zeta_{c,k}|^2} = \frac{\bs\zeta_{c,k} \wedge \bs{Q} \bs\zeta_{c,k}}{|\bs\zeta_{c,k}|^2} \tilde{\bs{\jmath}} = \beta_k \tilde{\bs{\jmath}} \,,
\end{equation}
where we have used $\tilde{\bs{\jmath}} = \bs{e}_1 \wedge \bs{e}_2 $ and re-obtained the real and imaginary parts of the eigenvalue $\bar\lambda_{r+2k-1}$ of the initial matrix $\A$. We observe thus that the definition of complex frequency coincides also in this case with the eigenvalue of the system, namely:
\begin{equation}
  \rho + \bs{\tilde{\omega}} = \alpha_k + \beta_k \tilde{\bs{\jmath}} \,,
\end{equation}
We note that the conjugate value of the eigenvalue can be obtained by defining:
\begin{equation}
  \bs{\zeta}' = \bs{D \zeta} - \bs{Q \zeta}\,.
\end{equation}
This changes the sign of $\bs{\tilde{\omega}}$, but not that of $\rho$ and simply indicates that the sign of the rotation (but not its magnitude) depends on the choice of the positive direction for the angles. 

\subsubsection*{Remark on the transformation matrix $\bs W$}

So far, we have shown that for an \ac{lti} system of order $n > 2$, after applying the transformation defined in \eqref{eq:transformation}, the complex frequencies of the decoupled subsystems of the transformed system coincide with the eigenvalues of the original system.  However, as in general $\bs W$ is not a rotation matrix, namely, a matrix with unit determinant and a matrix for which the inverse and the transpose coincide, the transformation from $\bs A$ to $\bs G$ is not isometric.  This means that, in general, even for second order linear ODEs, the complex frequency of the system before the transformation does not coincide with the eigenvalues of the state matrix.

\subsection{Linear systems with a defective state matrix}

For completeness, we briefly discuss the case of a system in the form of \eqref{eq:linear_system} with a defective state matrix.  
Since $\A$ is now non-diagonalizable, instead of the matrix $\bar{\bs{\Lambda}}$, $\A$ can be decomposed into $\bar{\bs{J}}$, namely the standard Jordan canonical form of $\A$.  Accordingly, the form of matrix $\bs{G}$ will change accordingly to include ones in its superdiagonal.

In the general case of an $n \times n$ defective system with a repeated eigenvalue $\bar\lambda$, the highest-order coupled generalized mode in the transformed coordinates $\bar \xi$ is given by:
\begin{equation}
     \bar \xi = \left( \sum_{k=0}^{n-1} \frac{c_{n-1-k}}{k!} t^k \right) e^{\bar\lambda t} \, , 
\end{equation}
where $c_{n-1-k}$ is the projection of the initial conditions onto the (n-1-k)-th mode.

We can then rewrite this sum as a standard polynomial $P$ of degree $n-1$:
\begin{equation}
    \bar \xi = P(t) \, e^{\bar\lambda t}\, , 
\end{equation}
where $P = m_{n-1}t^{n-1} + m_{n-2}t^{n-2} + \dots + m_1 t + m_0$, and $m_{n-1} \neq 0$.
Then:
\begin{equation}
    \frac{\bar \xi'}{\bar \xi} = \frac{P'e^{\bar\lambda t} + \bar \lambda Pe^{\bar\lambda t}}{Pe^{\bar \lambda t}} = \bar\lambda + \frac{P'}{P} \, .
\end{equation}
Because $P$ is a polynomial of degree $n-1$, its derivative $P'$ is a polynomial of degree $n-2$. Thus, as $t\to\infty$:
\begin{equation}
    \frac{\bar \xi'}{\bar \xi} \to \bar\lambda + 0 = \bar\lambda \, .
\end{equation}
Thus, in the case of a defective system, even though the geometric frequency does not at all times coincide with the eigenvalue, it always converges to it as $t \to \infty$.

\subsection{Nonlinear systems}

Consider a system of nonlinear ordinary differential equations (ODEs):
\begin{equation}
    \label{eq:nonlinear}
    \bs{x}' = \bs{f}(\bs{x}) \, ,
\end{equation}
Then, the time derivative of \eqref{eq:nonlinear} gives:
\begin{equation}
    \label{eq:dt_nonlinear}
    \begin{aligned}
    \bs{x}'' &= \nabla_{\bs{x}} \bs{f} \, \bs{x}' \\
    &= \A(\bs{x}) \, \bs{x}' \, ,
    \end{aligned}
\end{equation}
where $\A(\bs{x}) = \nabla_{\bs{x}} \bs{f}$ is the Jacobian matrix of the system, that during transients is time-varying.  
Equation \eqref{eq:nonlinear} can be written equivalently as:
\begin{equation}
    \label{eq:dt_nonlinear3}
    \bs{u}' = \A(\bs{x}) \, \bs{u} \, ,
\end{equation}
where $\bs{u}$ is the vector of generalized velocities.

As in general, $\A(\x)$ is time-varying during transients, the system cannot be decoupled into one- or two-order systems, as in this case $\bs{W}$ is also time-varying. In fact, if one defines $\bs{\zeta} = \bs{W} \bs{u}$, for a time-varying $\A = \bs{W}^{-1}{\bs G} {\bs W}$, we have:
\begin{equation}
  \bs{\zeta}' = \bs{W} \bs{u}' + \bs{W}' \bs{u}\,,
\end{equation}
and hence the diagonalization of $\A$ does not lead to a decoupled system:
\begin{equation}
  \bs{\zeta}' = [\bs{G} + \bs{W}'{\bs W}^{-1}] \bs{\zeta} \,,
\end{equation}
as matrix $\bs{W}' \bs{W}^{-1}$is not, in general, diagonal or block-diagonal as matrix $\bs{G}$.

We can however still apply the concepts of geometric frequency directly to the state trajectories of the system to study its dynamic behavior. 

Specifically when the system approaches a stable equilibrium we can conclude the following.
Assuming $\A$ is diagonalizable, it possesses $n$ eigenvalues $\bar\lambda_j$ with corresponding right eigenvectors $\bs{r}_j$, and left eigenvectors $\bs{l}_j$. The generalized velocities vector can be expressed as linear combination of its eigen-components:
\begin{equation}
\label{eq:exponentials}
    \bs{u} = \sum_{j=1}^nc_je^{\bar\lambda_jt}\bs{r}_j \, , 
\end{equation}
where the initial participation factors are $c_j = \bs{l}_i^{\bs{\top}} \bs{u}(0)$. To evaluate the asymptotic behavior as $t \rightarrow \infty$, we analyze the projection of $\bs{u}$ onto the manifold of the dominant mode.

Let the dominant eigenvalue $\bar\lambda_1$ to be real such that $\mu_1 <0$ and $\mu_1 > \mathrm{Re} (\bar\lambda_j)$ for all $j>1$.
\begin{equation}
\label{eq:dominant_exponentials}
    \bs{u} = e^{\mu_1 t} \left( c_1 \bs{r}_1 + \sum_{j=2}^n c_j e^{(\bar\lambda_j - \mu_1) t} \bs{r}_j \right) \, .
\end{equation}
with $\bs{r}_1$ and $\bs{r}_j$ orthonormal vectors.  Because $\mathrm{Re} (\bar\lambda_j) - \mu_1 < 0$, the summation terms exponentially decay and as $t$ increases:
\begin{equation}
\begin{aligned}
    \bs{u} &\rightarrow c_1 e^{\mu_1 t} \bs{r}_1 \, , \\
    \bs{u}' &\rightarrow 
    \mu_1 (c_1 e^{\mu_1 t} \bs{r}_1) = \mu_1 \bs{u} \, ,
\end{aligned}
\end{equation}
where we have utilized the fact that, close to equilibrium, $\bs{r}_1$ is constant.
Thus, $\rho$ and $\bs{\tilde{\omega}}$ will tend to:
\begin{equation}
\label{eq:real_dominant_eigenvalue}
\begin{aligned}   
  \rho &\rightarrow 
  \frac{\bs{u ^\top}(\mu_1 \bs{u})}{|\bs{u}|^2} = \mu_1  \frac{\bs{u ^\top}\bs{u}}{|\bs{u}|^2}= \mu_1 \,, \\
  |\bs{\tilde{\omega}}| &\rightarrow 
  \frac{|\bs{u}\wedge(\mu_1 \bs{u})|}{|\bs{u}|^2} = \mu_1  \frac{|\bs{u}\wedge\bs{u}|}{|\bs{u}|^2}= 0 \, .
\end{aligned}
\end{equation}
We can conclude that when the dominant mode of the linearized system approaching equilibrium is real, the trajectory of the generalized velocities collapses onto a 1D straight line. The symmetric part of the geometric frequency $\rho$ converges to the eigenvalue corresponding to that mode and because the wedge product of parallel vectors is zero,  $|\bs{\tilde{\omega}}|$ converges to zero as well. 

Now, let the dominant eigenvalues to be a complex conjugate pair $\bar\lambda_{1,2} = \alpha \pm \beta \jj$, where $\alpha <0$ and $\alpha > \text{Re}(\bar\lambda_j)$ for all $j > 2$.  For sufficiently high $t$, the velocity vector of the system will converge to:
\begin{equation}
    \label{eq:dominant_oscillatory_mode}
    \begin{aligned}
    \bs{u} &\rightarrow c_1 e^{\alpha t} \cos(\beta t) \bs{r}_1 + c_2 e^{\alpha t} \sin(\beta t) \bs{r}_2 \\
    &= e^{\alpha t} [\eta(t) \bs{r}_1 + \sigma(t) \bs{r}_2] \, ,    
    \end{aligned}
\end{equation}
with $|\bs{r_1}| = |\bs{r}_2| = 1$ and $\bs{r}_1 \perp \bs{r}_2$, that is, orthonormal vectors.  For $t$ sufficiently high, the time derivative of $\bs u$ tends to:
\begin{equation}
    \begin{aligned}
    \bs{u}' \rightarrow& \; c_1 e^{\alpha t} [\alpha  \cos(\beta t) - \beta \sin(\beta t)] \bs{r}_1 + \\
    & \; c_2 e^{\alpha t} [\alpha  \sin(\beta t) + \beta \cos(\beta t)] \bs{r}_2 \, ,        
    \end{aligned}
\end{equation}
or, equivalently:
\begin{equation}
    \begin{aligned}
    \bs{u}' \rightarrow& \; e^{\alpha t}[\alpha \eta(t) - c_{21} \beta \sigma(t)] \bs{r}_1 + \\
    & \; e^{\alpha t}[\alpha \sigma(t) + c_{12} \beta \eta(t)] \bs{r}_2 \, ,        
    \end{aligned}
\end{equation}
where $c_{21} = c_2/c_1$ and $c_{12} = c_1/c_2$.  Then, we obtain:
\begin{equation}
  \label{eq:rho_elliptical}
  \rho \rightarrow
  \alpha + \beta \, (c_{12} - c_{21}) \, \frac{ \eta(t)\sigma(t)}{\eta^2(t) + \sigma^2(t)} \, , 
\end{equation}
and
\begin{equation}
  \label{eq:omega_elliptical}
  |\tilde{\bs{\omega}}| \rightarrow \beta \, \frac{c_{12} \eta^2(t) + c_{21}\sigma^2(t)}{\eta^2(t) + \sigma^2(t)}  \, .
\end{equation}

As, in general, $c_1 \ne c_2$, $\rho$ and  $|\bs{\tilde{\omega}}|$ will oscillate around the values $\alpha$ and $\beta$, respectively.  Note that for $\beta=0$, we obtain the result for a real dominant eigenvalue given in \eqref{eq:real_dominant_eigenvalue}.  Moreover, for isotropic systems, i.e., for $c_1 = c_2$ and hence $c_{12}= c_{21} = 1$, we obtain $\rho \rightarrow \alpha$ and $|\tilde{\bs{\omega}}| \rightarrow \beta$.  This conditions imply that the state matrix has the structure of \eqref{eq:isomorphism}, and thus is equivalent to a complex number.  Thus, $\alpha$ and $\beta$ represents the complex frequency of the system where the states undergo a non-isotropic transformation, e.g., one of the state is ``stretched'' by a factor $c_{12}$.  In Section \ref{sec:nonlinear_circuit} below, we illustrates the cases for $\beta = 0$, $c_1 \ne c_2$ and $c_1 = c_2$.

\section{Examples}
\label{sec:examples}

The examples presented below apply the theory developed in the previous section and discuss applications to both linear and nonlinear system and to power systems. In particular, three cases of \ac{lti} circuits are discussed, namely, a first order RC circuit; a second order RLC circuit; and third order circuit consisted of the previous two connected in parallel.  Then, a nonlinear system, namely, a second-order tunnel-diode circuit, is analyzed in detail considering various scenarios, including a dominant monotonic mode, a dominant oscillatory mode, a dominant isotropic oscillatory mode, a system with multiple equilibria, and a stable limit cycle.

\subsection{RC Circuit}
\label{sec:RC_circuit}

\begin{figure}[h!]
  \centering
  \includegraphics[width=1.9in]{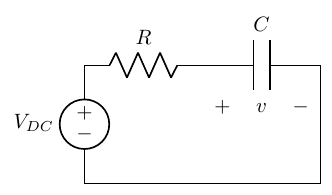}
  \caption{1st order RC circuit.}
  \label{fig:RC_circuit}
  \vspace{-2mm}
\end{figure}

Consider an RC circuit with a constant dc source $V_{DC}$ as can be seen in Fig. \ref{fig:RC_circuit}.
The differential equation that describes the system is:
\begin{equation}
   RC v' = - v \, + V_{DC} \,, 
\end{equation}
where $v \in \mathbb{R}$ is the voltage at the terminal of the capacitor and is a generalized position $\bs{x}$. By differentiating with respect to time, we get:
\begin{equation}
    v'' = - \frac{1}{RC}v' \,, 
\end{equation}
where $v'$ is a generalized velocity $\bs{u}$. The system is in the same form as \eqref{eq:system_speed} and its eigenvalue is $\bar\lambda = - \frac{1}{RC}$.
The complex frequency of $v'$ is defined using \eqref{eq:geom_freq}:
\begin{equation}
  \rho_{v'} + \bs{\tilde{\omega}}_{v'} = \frac{\bs{v'} \bs{v}''}{|v'|^2} = \frac{v' v''}{v'^2} = \frac{v''}{v'} = - \frac{1}{RC}  = \rho_{v'} \, ,
\end{equation}
which confirms that the eigenvalue of the system is also its complex frequency. As the system is one-dimensional, the complex frequency contains only the symmetric term $\rho_{v'}$.

\subsection{RLC Circuit}

\begin{figure}[h]
  \centering
  \includegraphics[width=2.5in]{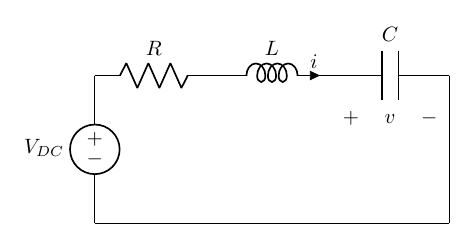}
  \caption{2nd order RLC circuit.}
  \label{fig:RLC_circuit}
  \vspace{-2mm}
\end{figure}

The series RLC circuit fed with a dc voltage shown in Fig. \ref{fig:RLC_circuit} is an example of a second-order \ac{lti} dynamic system. 

The ODEs that describe the dynamic behavior of the system are:
\begin{equation}
    \begin{bmatrix}
        i' \\ v'
    \end{bmatrix} = \begin{bmatrix}
        -\frac{R}{L} &  -\frac{1}{L} \\ 
        \frac{1}{C} &   0
    \end{bmatrix}\begin{bmatrix}
        i \\ v
    \end{bmatrix} + \begin{bmatrix}
        \frac{1}{L} \\ 0
    \end{bmatrix} V_{DC} = \A \begin{bmatrix}
        i \\ v
    \end{bmatrix} + \begin{bmatrix}
        \frac{1}{L} \\ 0
    \end{bmatrix} V_{DC} \, .
\end{equation}
By differentiating with respect to time, we get:
\begin{equation}
    \begin{bmatrix}
        i'' \\ v''
    \end{bmatrix} = \begin{bmatrix}
        -\frac{R}{L} &  -\frac{1}{L} \\ 
        \frac{1}{C} &   0
    \end{bmatrix}\begin{bmatrix}
        i' \\ v'
    \end{bmatrix} = \A \begin{bmatrix}
        i' \\ v'
    \end{bmatrix} \, .
\end{equation}
The system is in the form of \eqref{eq:system_speed} and, solving ${\rm det}(\bs{A}-\lambda \bs{I})=0$, we find the eigenvalues of the system:
\begin{equation}
    \bar{\lambda}_{1,2} = \alpha + \beta\jj=  -\frac{R}{2L}\pm \sqrt{ \left( \frac{R}{2L}\right) ^2-\frac{1}{LC}} \,.
\end{equation}
In this example, our goal is to examine the case where a second-order \ac{lti} system has a pair of complex conjugate eigenvalues. For the system to have a pair of complex conjugate eigenvalues the following relationship has to be true:
\begin{equation*}
    \left( \frac{R}{2L}\right)^2-\frac{1}{LC} < 0 \Rightarrow R^2 < \frac{4L}{C} \,.
\end{equation*}
By applying the procedure described in Section \ref{sec:connection} for systems where $\A$ is a 2-by-2 time-invariant matrix, the transformed system becomes:
\begin{equation}
\begin{aligned}
        \bs{\zeta}' &= \bs{G}\bs{\zeta} = 
        \begin{bmatrix}
        \alpha & -\beta \\ 
        \beta  & \alpha 
    \end{bmatrix} 
    \bs{\zeta} 
    \\ 
     &= \begin{bmatrix}
           -\frac{R}{2L} & -\sqrt{(\frac{R}{2L})^2-\frac{1}{LC}} \\ 
          \sqrt{(\frac{R}{2L})^2-\frac{1}{LC}}  & -\frac{R}{2L}
      \end{bmatrix}
    \begin{bmatrix}
        \zeta_1 \\ \zeta_2 
    \end{bmatrix} \,.
\end{aligned}
\end{equation}
Matrix $\bs{G}$ can be decomposed into a symmetric and an anti-symmetric part:
\begin{equation}
    \bs{\zeta}' = \bs{D}\bs{\zeta} + \bs{Q}\bs{\zeta},
\end{equation}
where:
\begin{equation}
    \bs{D}  =
    \begin{bmatrix}
        -\frac{R}{2L} & 0 \\ 
        0  & -\frac{R}{2L} 
    \end{bmatrix} = \begin{bmatrix}
        \alpha & 0 \\ 
        0  & \alpha 
    \end{bmatrix} , \
\end{equation}
and:
\begin{equation}   
    \bs{Q} =
    \begin{bmatrix}
        0 & -\sqrt{(\frac{R}{2L})^2-\frac{1}{LC}} \\ 
        \sqrt{(\frac{R}{2L})^2-\frac{1}{LC}}  & 0
    \end{bmatrix} = \begin{bmatrix}
        0 & -\beta \\ 
        \beta  & 0 
    \end{bmatrix} .
\end{equation}
Then the real part of the complex frequency of the system is:
\begin{equation}
\label{eq:2nd_order_rho}
\begin{aligned}
    \rho = \frac{\bs{\zeta} \cdot \bs{\zeta}'}{|\bs{\zeta}|^2} 
    = & \frac{\bs{\zeta^\top} \bs{D}\bs{\zeta}}{|\bs{\zeta}|^2} \\ 
    = & \frac{\begin{bmatrix}
        \zeta_1 & \zeta_2
    \end{bmatrix} \begin{bmatrix}
        -\frac{R}{2L} & 0 \\  0 &  -\frac{R}{2L}
    \end{bmatrix} \begin{bmatrix}
        \zeta_1  \\  \zeta_2 
    \end{bmatrix}}{\den{1}{2}}  \\ 
    = & 
    \frac{\begin{bmatrix}
        -\frac{R}{2L}\zeta_1 & -\frac{R}{2L}\zeta_2
    \end{bmatrix}  \begin{bmatrix}
        \zeta_1  \\  \zeta_2 
    \end{bmatrix}}{\den{1}{2}} 
    = \frac{
        -\frac{R}{2L}\zeta_1^2 -\frac{R}{2L}\zeta_2^2
    }{\den{1}{2}} \\ 
    = & \frac{
        -\frac{R}{2L}(\zeta_1^2 + \zeta_2^2)
    }{\den{1}{2}} 
    =  -\frac{R}{2L} = \alpha \, .
\end{aligned}
\end{equation}
Similarly, the imaginary part of the complex frequency is: 
\begin{equation}
\label{eq:2nd_order_omega}
\begin{aligned}
    \bs{\tilde{\omega}}=\frac{\bs{\zeta} \wedge \bs{\zeta}'}{\den{1}{2}} = & \frac{\bs{\zeta} \wedge \bs{Q}\bs{\zeta}}{\den{1}{2}} \\ 
    = & \frac{(\zeta_1 \zeta_2' - \zeta_2 \zeta_1')}{\den{1}{2}}(\bs{e}_1 \wedge \bs{e}_2) \\ 
    =&  \frac{(\zeta_1^2 +\zeta_2^2)\sqrt{\left (\frac{R}{2L} \right )^2-\frac{1}{LC}}}{\den{1}{2}}\jj \\ 
    =&  \sqrt{\left (\frac{R}{2L} \right )^2-\frac{1}{LC}} \, \jj = \beta \jj \, .
\end{aligned}
\end{equation}

Equations \eqref{eq:2nd_order_rho} and \eqref{eq:2nd_order_omega} show that the real and imaginary components of the complex frequency computed from the system states, subject to the non-isometric transformation defined in Section \ref{sec:connection}, coincides with the real and imaginary parts, respectively, of the eigenvalues of the original system.

\subsection{Third-order System}

\begin{figure}[htb]
  \centering
  \includegraphics[width=3.2in]{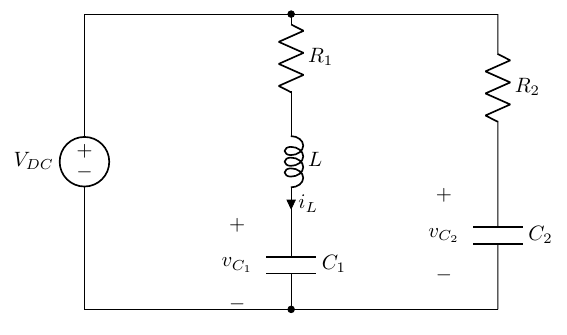}
  \caption{3rd order circuit.}
  \label{fig:3rd_Order_Circuit}
  \vspace{-2mm}
\end{figure}

In this example, we illustrate the application of our methodology in a third-order \ac{lti} system.  The system is comprised of a series RLC branch and a series RC branch connected in parallel across a voltage source, as can be seen in Fig. \ref{fig:3rd_Order_Circuit}. The purpose of this example is to demonstrate the proposed methodology for systems of order greater than two, where each decoupled subsystem is studied independently. The system matrix $\A$ has one real eigenvalue and one pair of complex-conjugate eigenvalues, so that both cases corresponding to $n \leq 2$ appear.

The system of ODEs that describe this circuit is:
\begin{equation}
    \begin{bmatrix}
        v_{C1}' \\i_L' \\ v_{C2}'
    \end{bmatrix} = \begin{bmatrix}
        0 &  2 & 0 \\ 
        -1 &  -1 & 0 \\ 
        0 &  0 & -1 
    \end{bmatrix}\begin{bmatrix}
        v_{C1} \\i_L \\ v_{C2}
    \end{bmatrix} + \begin{bmatrix}
        0 \\1 \\ 1
    \end{bmatrix} V_{DC}\, .
\end{equation}
By differentiating the original system with respect to time, the system becomes of the form \eqref{eq:system_speed}:
\begin{equation}
    \begin{bmatrix}
        v_{C1}'' \\i_L'' \\ v_{C2}''
    \end{bmatrix} = \begin{bmatrix}
        0 &  2 & 0 \\ 
        -1 &  -1 & 0 \\ 
        0 &  0 & -1 
    \end{bmatrix}\begin{bmatrix}
        v_{C1}' \\i_L' \\ v_{C2}'
    \end{bmatrix} = \A\begin{bmatrix}
        v_{C1}' \\i_L' \\ v_{C2}'
    \end{bmatrix} \, ,
\end{equation}
This state matrix arises from the following values of the system's parameters: 
\begin{itemize}
    \item Series RLC: $R_1 =  1\, \Omega$, $L = 1.0 \, H$, $C_1 = 0.5 \, C$ . 
    \item Series RC: $R_2 = 1 \, C$, $C_2 = 1 \, C$ .
\end{itemize}

The eigenvalues of the system matrix $\bs{A}$ are:
\begin{equation}
    \bar\lambda_1 = -1 \,,  \qquad\bar\lambda_{2,3} = \frac{-1 \pm \sqrt{7}\jj}{2}  \,.
\end{equation}

Since $\A$ is diagonalizable, by applying the transformation defined in \eqref{eq:transformation}, utilizing the decomposition of $\A$ defined in \eqref{eq:decomposition_matrix_real}, we get:
\begin{equation}
    \bs{G} =
    \begin{bmatrix}
    -1 & 0 & 0 \\ 
    0 & -1/2 & -\sqrt{7}/2 \\ 
    0 & \sqrt{7}/2 & -1/2
    \end{bmatrix} \,,
\end{equation}
where the two decoupled system become now obvious.

The first of the two decoupled systems is:
\begin{equation}
    \zeta_1' = -\zeta_1 \,.
\end{equation}
The real part of the complex frequency of this system is:
\begin{equation}
\label{eq:3rd_order_1_rho}
    \rho = \frac{\zeta_1 \cdot \zeta'_1}{|\zeta_1|^2}  = \frac{-\zeta_1\zeta_1}{|\zeta_1|^2} = -1 \,,
\end{equation}
while the imaginary part of the complex frequency is zero as the eigenvalue is real.

The second decoupled system is:
\begin{equation}
    \begin{bmatrix} 
    \zeta'_2 \\ \zeta'_3
    \end{bmatrix} = \begin{bmatrix} 
    -1/2    &   -\sqrt{7}/2 \\ \sqrt{7}/2    &   -1/2
    \end{bmatrix}\begin{bmatrix} 
    \zeta_2 \\ \zeta_3
    \end{bmatrix} \,.
\end{equation}
The real part of the complex frequency of this system is:
\begin{equation}
\label{eq:3rd_order_2_rho}
\begin{aligned}
    \rho &= \frac{\begin{bmatrix} 
    \zeta_2 \\ \zeta_3
    \end{bmatrix}
    \cdot \begin{bmatrix} 
    \zeta'_2 \\ \zeta'_3
    \end{bmatrix}}{\den{2}{3}} = \frac{\begin{bmatrix} 
    \zeta_2 \\ \zeta_3
    \end{bmatrix}^{\bs{\top}} \begin{bmatrix} 
    -1/2    &   0 \\ 0    &   -1/2
    \end{bmatrix} \begin{bmatrix} 
    \zeta_2 \\ \zeta_3
    \end{bmatrix}}{\den{2}{3}} \\ 
    &= \frac{\begin{bmatrix} 
    \zeta_2 & \zeta_3
    \end{bmatrix}  \begin{bmatrix} 
    -1/2    &   0 \\ 0    &   -1/2
    \end{bmatrix} \begin{bmatrix} 
    \zeta_2 \\ \zeta_3
    \end{bmatrix}}{\den{2}{3}} \\ 
    &= \frac{\begin{bmatrix} 
    -(1/2) \zeta_2 & -(1/2) \zeta_3
    \end{bmatrix}  \begin{bmatrix} 
    \zeta_2 \\ \zeta_3
    \end{bmatrix}}{\den{2}{3}} \\ 
    &= \frac{-(1/2) (\zeta^2_2 + \zeta^2_3)}{\den{2}{3}} = -\frac{1}{2}  \,,
\end{aligned}
\end{equation}
whereas the imaginary part of the complex frequency is:
\begin{equation}
\label{eq:3rd_order_2_omega}
\begin{aligned}
\bs{\tilde{\omega}} &= \frac{\begin{bmatrix} 
    \zeta_2 \\ \zeta_3
    \end{bmatrix} \wedge \begin{bmatrix} 
    \zeta'_2 \\ \zeta'_3
    \end{bmatrix}}{\den{2}{3}} = \frac{\begin{bmatrix} 
    \zeta_2 \\ \zeta_3
    \end{bmatrix} \wedge \begin{bmatrix} 
    0   &  -\sqrt{7}/2 \\ -\sqrt{7}/2    &   0
    \end{bmatrix}\begin{bmatrix} 
    \zeta_2 \\ \zeta_3
    \end{bmatrix}}{\den{2}{3}}  \\ 
    &= \frac{\begin{bmatrix} 
    \zeta_2 \\ \zeta_3
    \end{bmatrix} \wedge \begin{bmatrix} 
    -(\sqrt{7}/2)\zeta_3 \\ -(\sqrt{7}/2)\zeta_2
    \end{bmatrix}}{\den{2}{3}} = \frac{-(\sqrt{7}/2) (\zeta^2_2 + \zeta^2_3)}{\den{2}{3}}\bs{e}_1 \wedge \bs{e}_2 \\ 
    &= -\frac{\sqrt{7}}{2} \jj\,.
\end{aligned}
\end{equation}

In summary, \eqref{eq:3rd_order_1_rho} shows that the real part of the complex frequency of the first subsystem is equal to the real part of its corresponding eigenvalue;  \eqref{eq:3rd_order_2_rho} shows that the real part of the complex frequency of the second subsystem is equal to the real part of its corresponding eigenvalue; and \eqref{eq:3rd_order_2_omega} shows that the imaginary part of the complex frequency is equal to the imaginary part of its corresponding eigenvalue.

\subsection{Second-order Nonlinear Circuit}
\label{sec:nonlinear_circuit}

\begin{figure}[htb]
  \centering
  \includegraphics[width=3.2in]{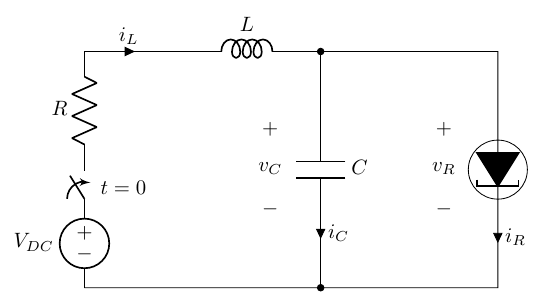}
  \caption{Tunnel diode circuit.}
  \label{fig:tunnel_diode}
\end{figure}

This section discusses an example of a $2 \times 2$ nonlinear circuit.  The circuit shown in Fig.~\ref{fig:tunnel_diode} represents a tunnel diode circuit and is taken from \cite{chua1987linear}.  As discussed in Section \ref{sec:connection}, the transformation that we defined for the \ac{lti} systems cannot be applied.  We instead focus on the complex frequency of the whole system and compare it with the eigenvalues of the system.  In the remainder of this section, we consider various scenarios: dominant monotonic mode, that is, the dominant eigenvalue at equilibrium is real; dominant oscillatory mode, that is, the dominant eigenvalues at equilibrium are complex conjugate; and a stable limit cycle.  A case with multiple equilibria is also discussed.

\subsubsection{Dominant Monotonic Mode} 

\begin{figure}[htb]
  \centering
  \includegraphics[width=0.99\linewidth]{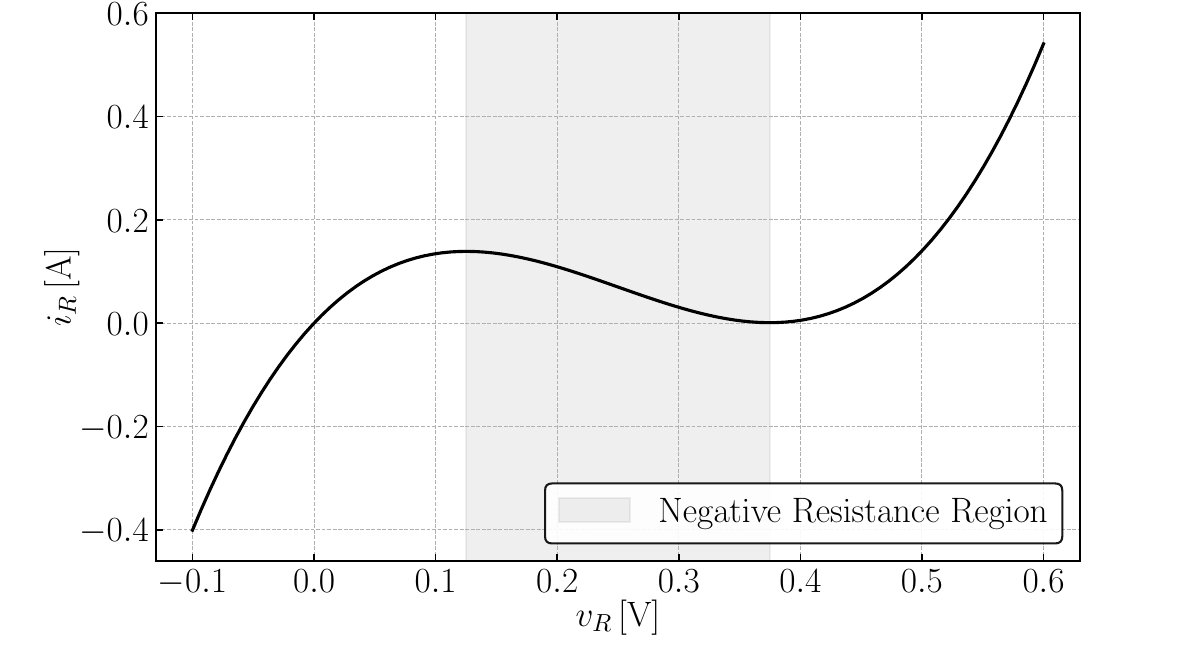}
  \caption{Tunnel diode $v_R-i_R$ characteristic.}
  \label{fig:tunnel_diode_char}
\end{figure}

The parameters of the system's components are the following:
\begin{itemize}
    \item Inductance: $L = 1.0 $ H
    \item Capacitance: $C = 0.5 $ F
    \item Resistance:  $R = 0.2 \ \Omega$
    \item Voltage Source: $V_{DC} = 0.5 $ V
\end{itemize}
The tunnel diode's $v-i$ characteristic can be seen in Fig. \ref{fig:tunnel_diode_char}. 
By choosing $v_{C}$ and $i_{L}$ as the state variables, the equations that define the dynamic system are:
\begin{equation}
    \label{eq:tunneldiodecircuit}
    \begin{bmatrix}
        v'_{C}  \\ 
        i'_{L}
    \end{bmatrix} =
    \begin{bmatrix}
        \frac{1}{C}(-i_R {\scriptstyle (v_{C})} + i_{L}) \\ 
        \frac{1}{L}(V_{DC}-Ri_{L} - v_{C})
    \end{bmatrix} \\ 
    \,.
\end{equation}
The time variable $t$ does not appear in the state equation \eqref{eq:tunneldiodecircuit} since the circuit contains only time-invariant elements and a dc source. 

\begin{figure}[htb]
  \centering
  \includegraphics[width=0.99\linewidth]{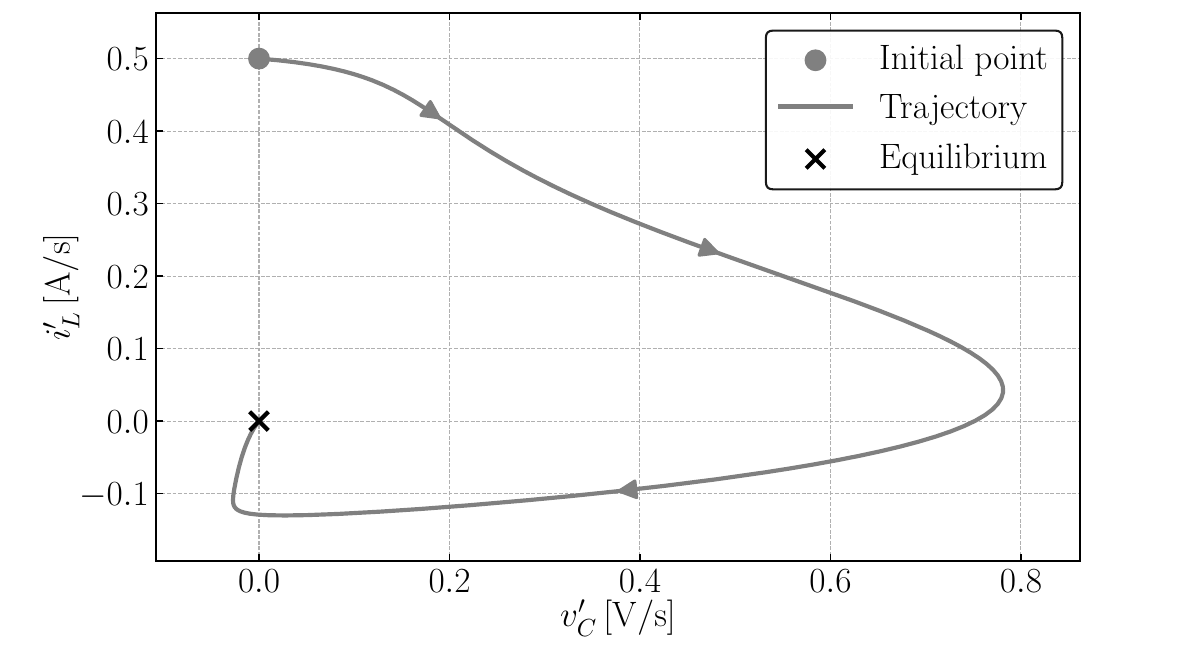}
  \caption{State space trajectory of generalized velocities when the tunnel diode system goes to a new equilibrium.}
  \label{fig:tunnel_diode_transformed_phase_u}
\end{figure}

The system in \eqref{eq:tunneldiodecircuit} is in the form of \eqref{eq:nonlinear}, thus to calculate the complex frequency of the system we first calculate its generalized velocity as defined in \eqref{eq:position_to_velocity}.  The trajectory of the system's generalized velocities $\bs{u}$ can be seen in Fig.~\ref{fig:tunnel_diode_transformed_phase_u}.

To observe how the value of the complex frequency changes compared to the eigenvalues during the time domain simulation, the eigenvalues of the linearized system at each time step of the simulation are calculated.

The system is initially not energized.  At $t = 0$ s, the switch closes and the battery is connected to the rest of the system. After some time the system reaches its state of equilibrium, as shown in Fig. \ref{fig:tunnel_diode_transformed_phase_u}.

\begin{figure}[htb]
  \centering
  \includegraphics[width=0.99\linewidth]{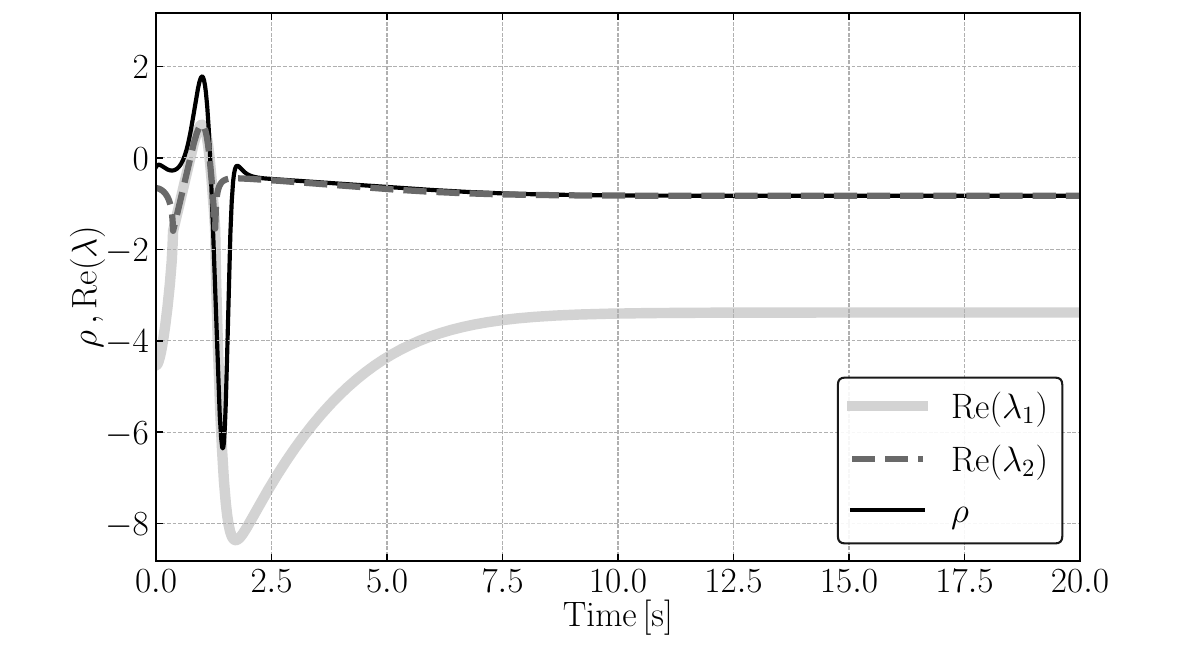}
  \caption{Real part of the complex frequency and of the eigenvalues of the tunnel diode system as it reaches a new equilibrium.}
  \label{fig:tunnel_diode_rho_vs_eigenvalues}
\end{figure}

Figure \ref{fig:tunnel_diode_rho_vs_eigenvalues} shows the real part of the complex frequency and the eigenvalues of the linearized system at each time step are shown.  As the system reaches its equilibrium, $\rho$ takes the same value as the system's dominant eigenvalue.

\begin{figure}[htb]
  \centering
  \includegraphics[width=0.99\linewidth]{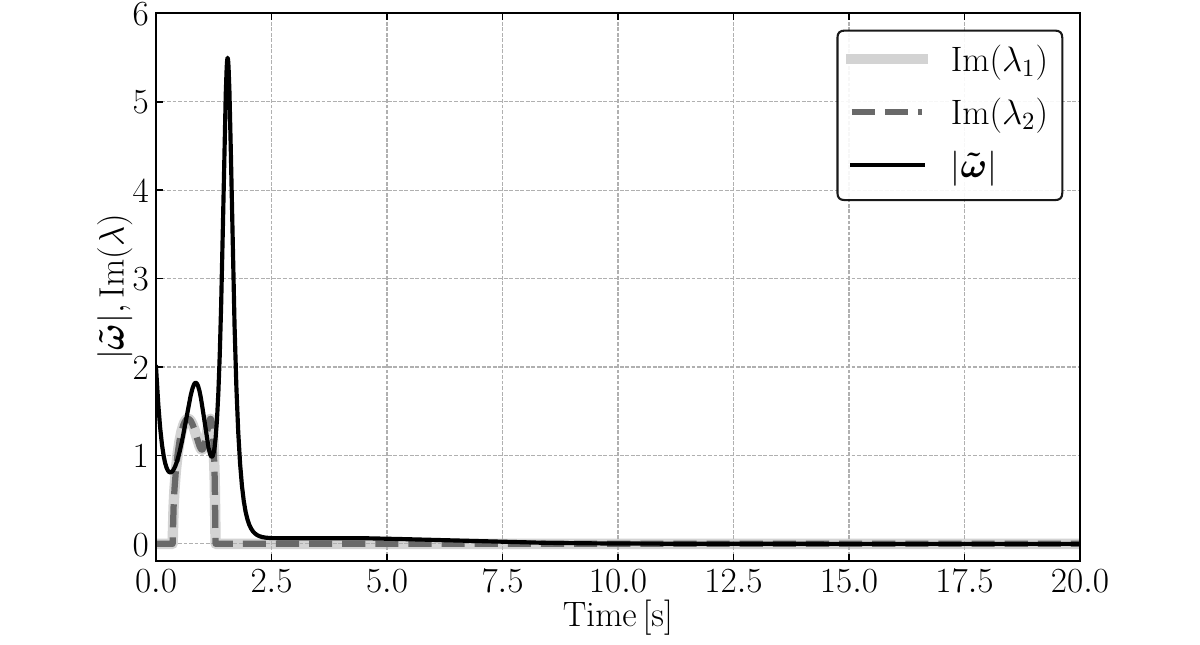}
  \caption{Imaginary part of complex frequency and of the eigenvalues of the tunnel diode system as it reaches a new equilibrium.}
  \label{fig:tunnel_diode_omega_vs_eigenvalues}
\end{figure}

In Fig.~\ref{fig:tunnel_diode_omega_vs_eigenvalues}, the imaginary part of the complex frequency and the eigenvalues of the linearized system at each time step are shown. $|\omegau{u}|$ is positive during the whole trajectory of the system and goes to zero at the equilibrium. At the same time, there is only a specific time period where the two eigenvalues become complex and their imaginary part exists.

\subsubsection{Dominant Oscillatory Mode}
\label{sec:tunnel_diode_extra}

\begin{figure}[htb]
  \centering
  \includegraphics[width=0.99\linewidth]{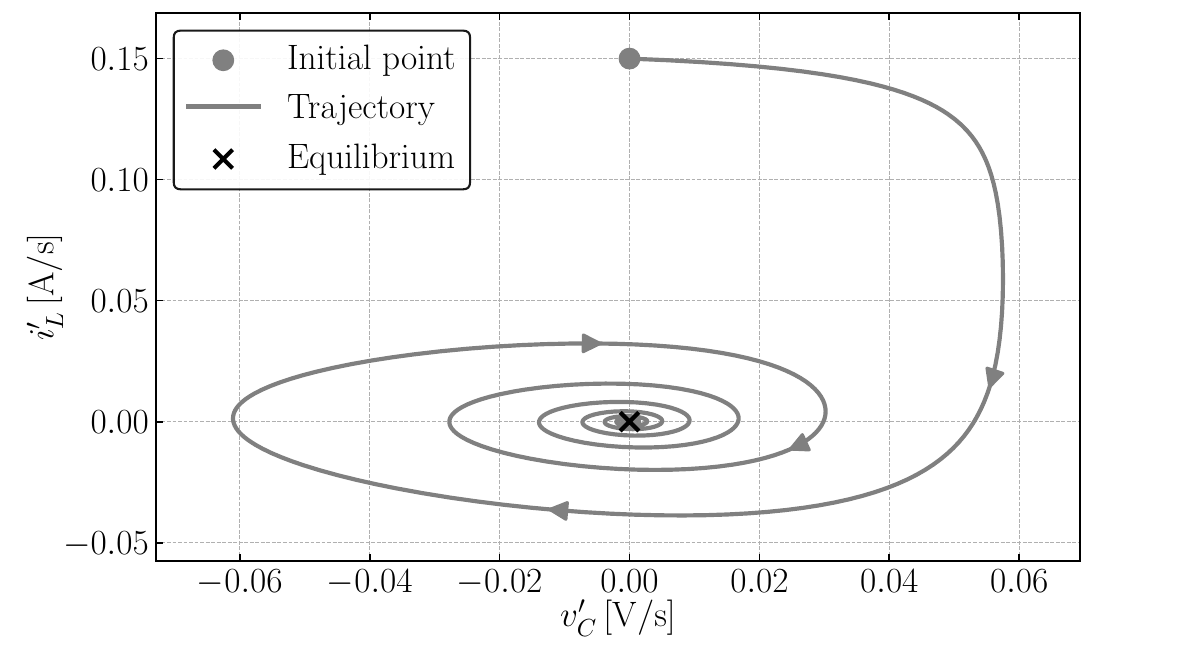}
  \caption{State space trajectory of generalized velocities when the tunnel diode's dominant mode is oscillatory.}
  \label{fig:tunnel_diode_phase_u_oscillatory}
  \vspace{-2mm}
\end{figure}

We consider again the tunnel diode circuit of Fig.~\ref{fig:tunnel_diode} and change the value of the voltage source to $V_{DC} = 0.15$~V.  The system is initially not energized.  At $t = 0$ s the switch closes.   Figure \ref{fig:tunnel_diode_phase_u_oscillatory} shows the state space trajectory of the generalized velocities of the circuit.  The new value of $V_{DC}$ causes the system to have an oscillatory dominant mode.

\begin{figure}[htb]
  \centering
  \includegraphics[width=0.99\linewidth]{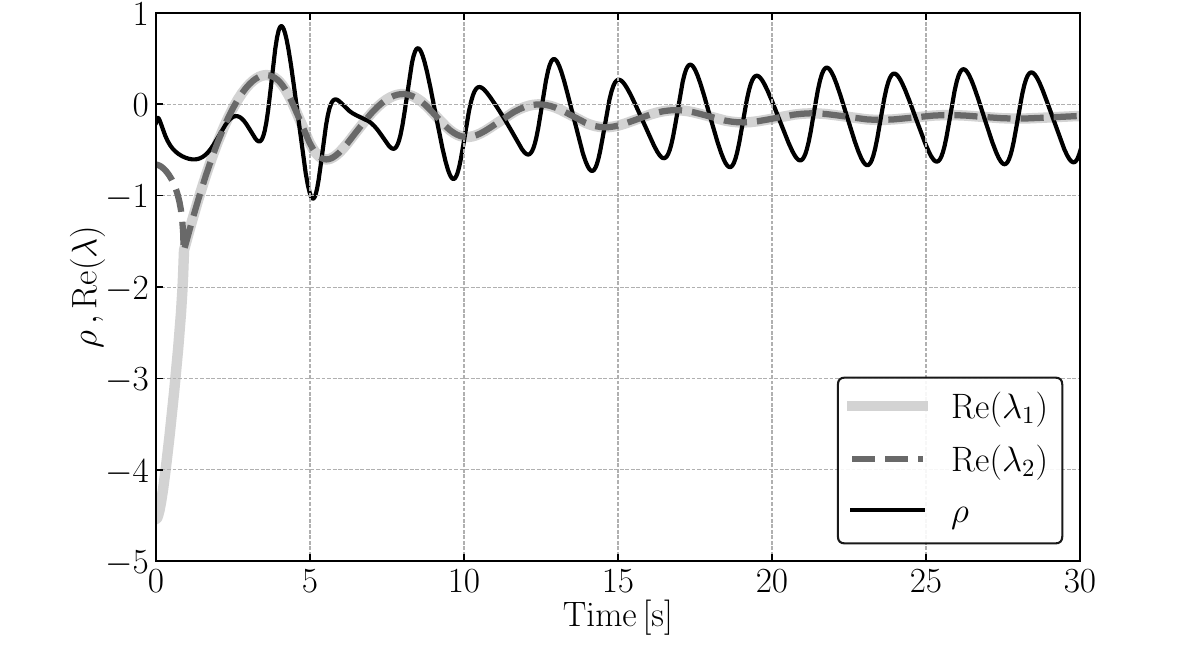}
  \caption{Real part of the complex frequency and of the eigenvalues of the tunnel diode system when the dominant mode of the system is oscillatory.}
  \label{fig:tunnel_diode_rho_vs_eigenvalues_oscillatory}
  \vspace{-2mm}
\end{figure}

\begin{figure}[htb]
  \centering
  \includegraphics[width=0.99\linewidth]{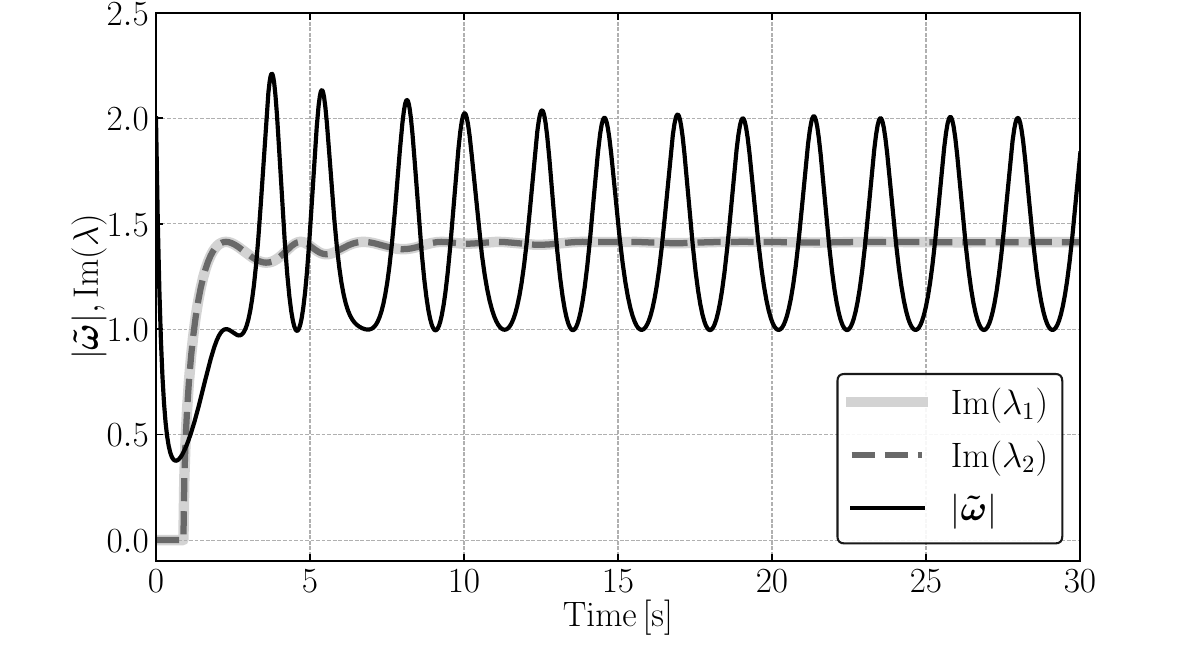}
  \caption{Imaginary part of the complex frequency and of the eigenvalues of the tunnel diode system when the dominant mode of the system is oscillatory.}
  \label{fig:tunnel_diode_omega_vs_eigenvalues_oscillatory}
  \vspace{-2mm}
\end{figure}

The transient behavior of $\rho$ vs the value of the real part of the eigenvalues of the system is shown in Fig.~\ref{fig:tunnel_diode_rho_vs_eigenvalues_oscillatory}, whereas  the imaginary part of the complex frequency and the imaginary part of the eigenvalues of the system are shown in Fig.~\ref{fig:tunnel_diode_omega_vs_eigenvalues_oscillatory}.    The system approaches the equilibrium with an exponentially decaying oscillation.  
As predicted with \eqref{eq:rho_elliptical} and \eqref{eq:omega_elliptical}, $\rho$ and $|\tilde{\bs{\omega}}|$ oscillate around the value of the real and imaginary parts, respectively, of the complex eigenvalue towards which the Jacobian matrix of the circuit converges as $t$ increases.  The eigenvalue of the Jacobian matrix, thus, represents the complex frequency of a non-isometric transformation of the system for which the trajectory of the transformed stated is characterized by $c_1 = c_2$ in equations \eqref{eq:rho_elliptical} and \eqref{eq:omega_elliptical}.

\subsubsection{Dominant Isotropic Oscillatory Mode}

In this scenario, we assume again a domain oscillatory mode but we choose the parameters in such a way that the mode is isotropic, namely $c_1 = c_2$ in \eqref{eq:dominant_oscillatory_mode}.  
We use the following parameters:
\begin{itemize}
    \item Inductance: $L = 1.0$ H
    \item Capacitance: $C = 1.0$ F
    \item Resistance:  $R = 0.3688 \ \Omega$
    \item Voltage Source: $V_{DC} = 0.402$ V
\end{itemize}

\begin{figure}[htb]
  \centering
  \includegraphics[width=0.99\linewidth]{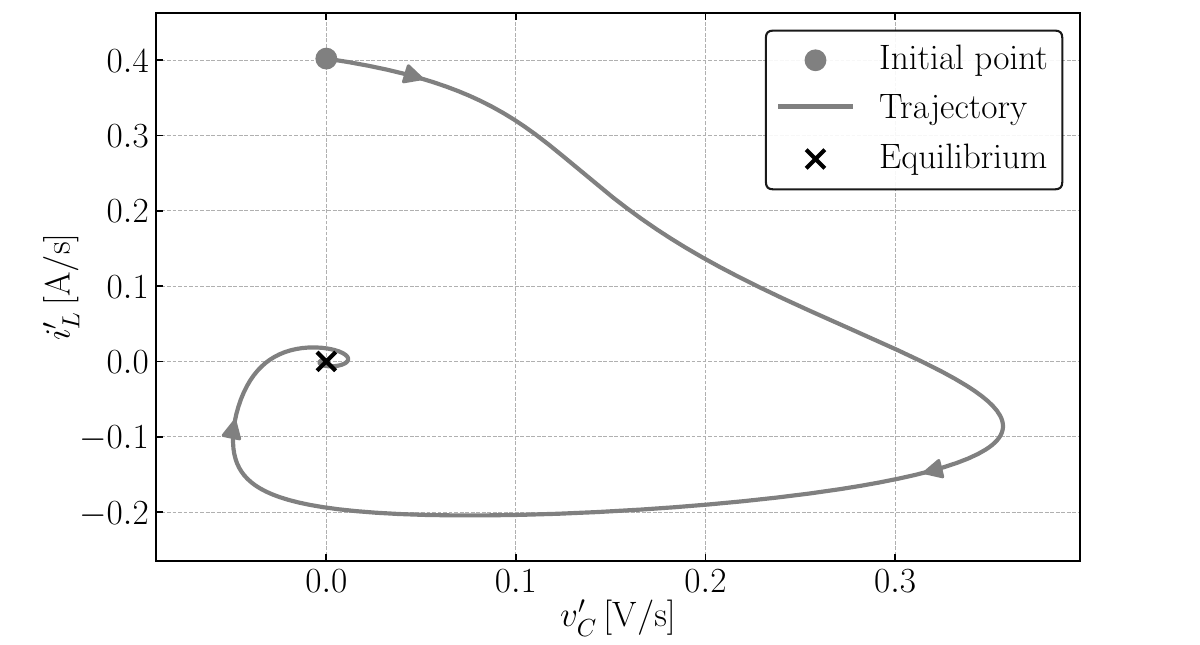}
  \caption{State space trajectory of generalized velocities of the tunnel diode for an isotropic oscillation.}
  \label{fig:tunnel_diode_phase_u_circle}
  \vspace{-2mm}
\end{figure}

\begin{figure}[htb]
  \centering
  \includegraphics[width=0.99\linewidth]{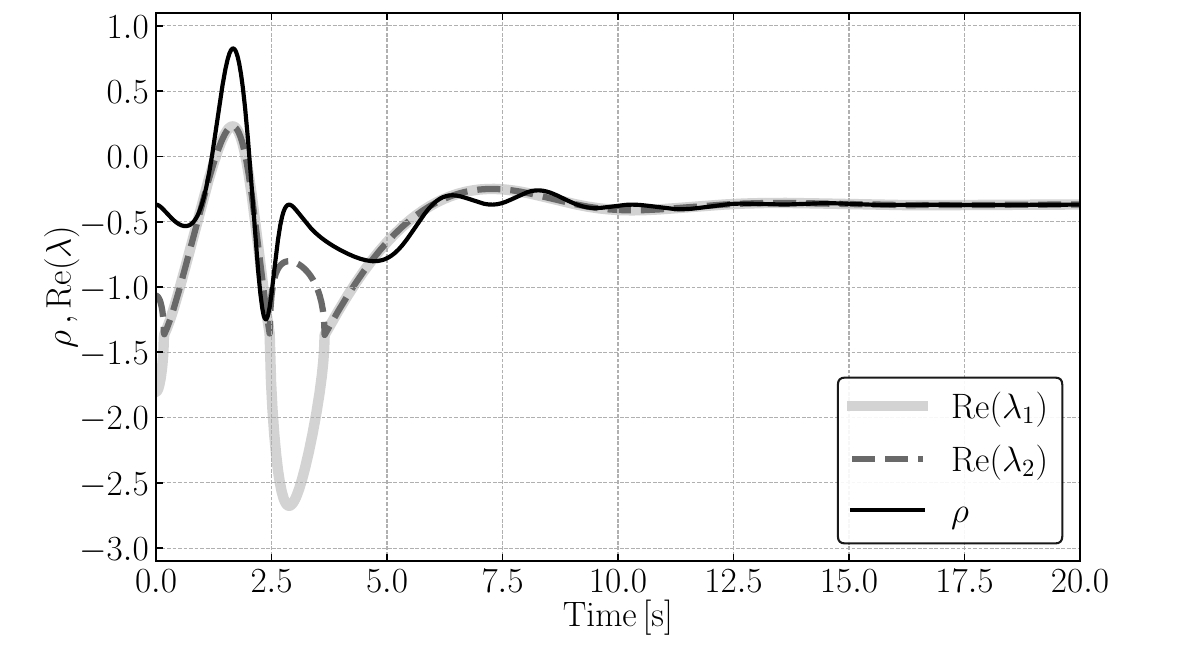}
  \caption{Real part of the complex frequency and of the eigenvalues of the tunnel diode system for an isotropic oscillation.}
  \label{fig:tunnel_diode_rho_vs_eigenvalues_circle}
  \vspace{-2mm}
\end{figure}

\begin{figure}[htb]
  \centering
  \includegraphics[width=0.99\linewidth]{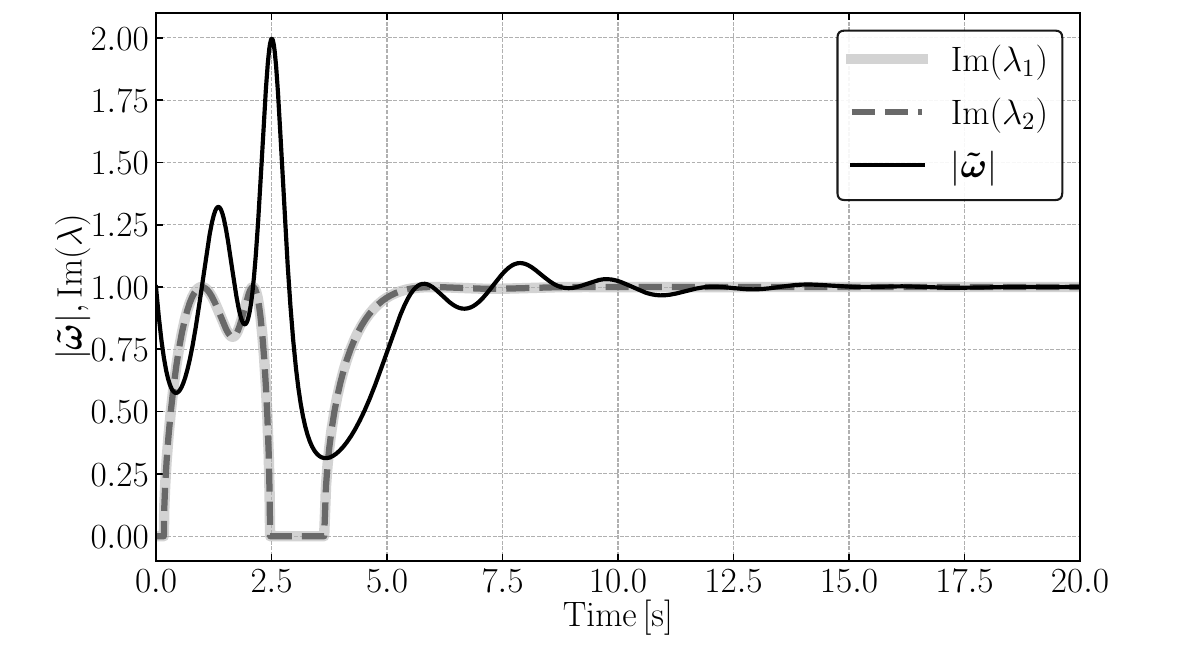}
  \caption{Imaginary part of the complex frequency and of the eigenvalues of the tunnel diode system for an isotropic oscillation.}
  \label{fig:tunnel_diode_omega_vs_eigenvalues_circle}
  \vspace{-2mm}
\end{figure}

Figure \ref{fig:tunnel_diode_phase_u_circle} shows the trajectory of the generalized velocities of the circuit, whereas Figs.~\ref{fig:tunnel_diode_rho_vs_eigenvalues_circle} and \ref{fig:tunnel_diode_omega_vs_eigenvalues_circle} show the transient of $\rho$ and $|\tilde{\bs{\omega}}|$ vs the real and imaginary parts, respectively, of the eigenvalues of the Jacobian matrix of the circuit.  


\subsubsection{Limit Cycle}

\begin{figure}[htb]
  \centering
  \includegraphics[width=0.99\linewidth]{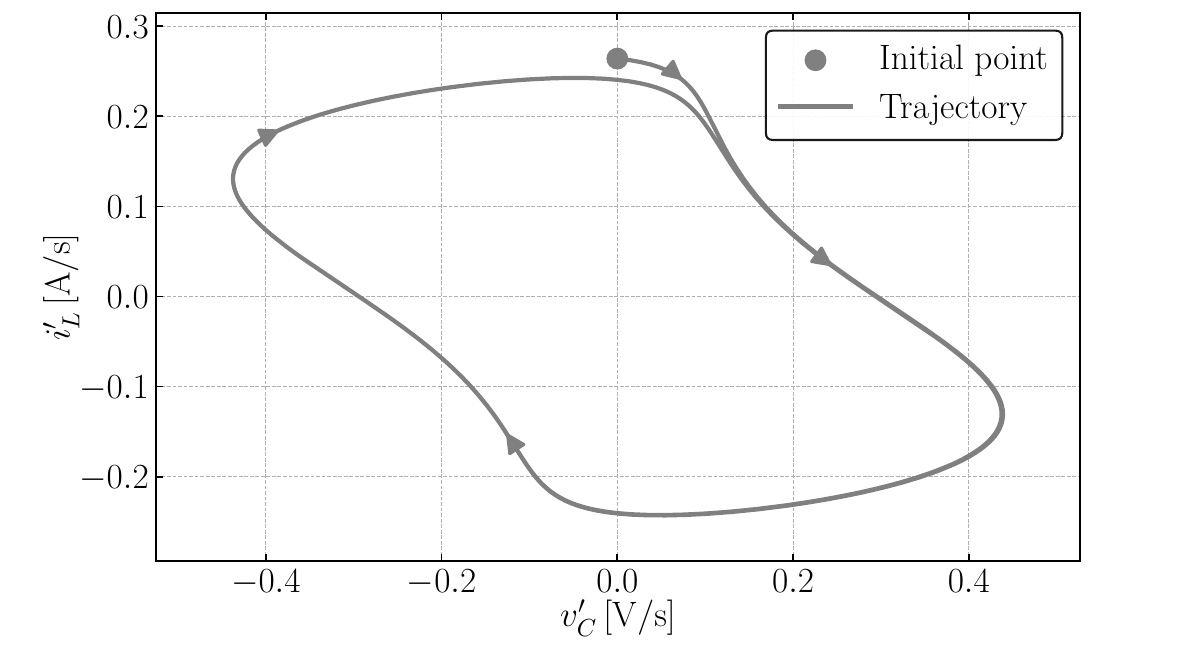}
  \caption{State space trajectory of generalized velocities when the tunnel diode system goes to a limit cycle.}
  \label{fig:tunnel_diode_phase_u_limit_cycle}
\end{figure}

We consider the same system with the only difference that now the voltage of the dc voltage source is $V_{DC} = 0.264$ V.
In this scenario, instead of the system reaching to a new equilibrium state after the closing of the switch, the system reaches a limit cycle as can be seen from Fig.~\ref{fig:tunnel_diode_phase_u_limit_cycle}.

\begin{figure}[htb]
  \centering
  \includegraphics[width=0.99\linewidth]{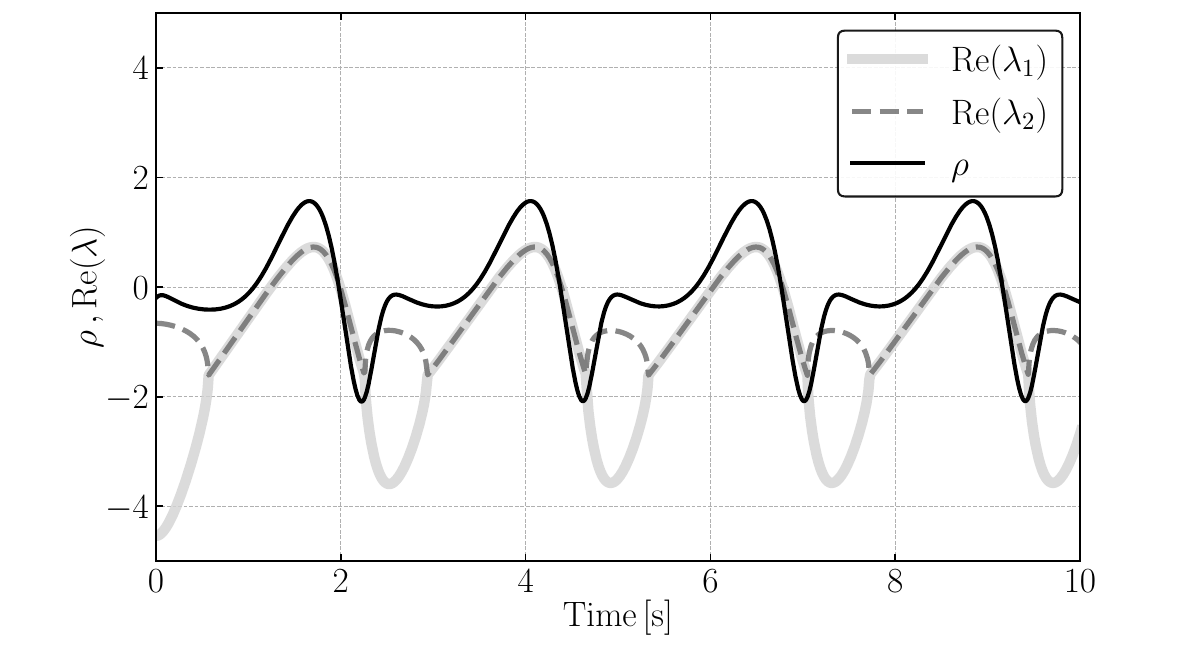}
  \caption{Real part of the complex frequency and of the eigenvalues of the tunnel diode system when it goes to a limit cycle.}
  \label{fig:tunnel_diode_rho_vs_eigenvalues_limit_cycle}
  \vspace{-2mm}
\end{figure}

Figure \ref{fig:tunnel_diode_rho_vs_eigenvalues_limit_cycle} shows the real part of the complex frequency and the eigenvalues of the linearized system at each time step. In some intervals, the gradient of the system flows has two real eigenvalues, and in others, it has a pair of complex conjugate ones.  $\rhou{u}$ is oscillating as the system never reaches to an equilibrium point.

\begin{figure}[t!]
  \centering
  \includegraphics[width=0.99\linewidth]{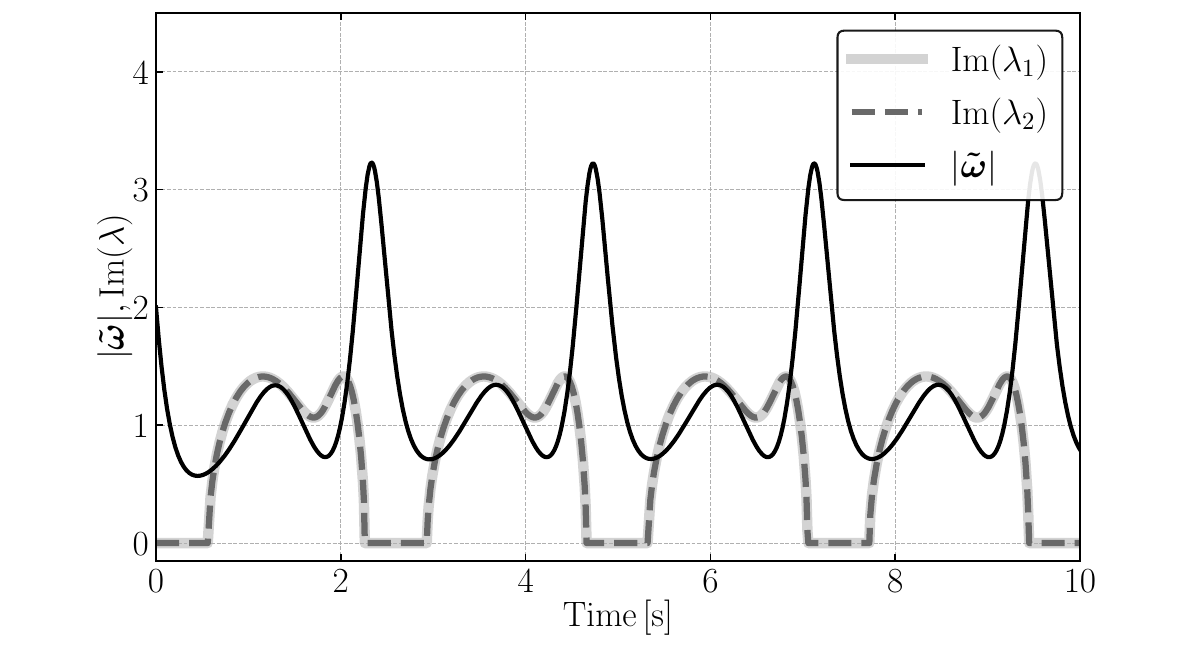}
  \caption{Imaginary part of complex frequency and of the eigenvalues of the tunnel diode system when it goes to a limit cycle.}
  \label{fig:tunnel_diode_omega_vs_eigenvalues_limit_cycle}
  \vspace{-2mm}
\end{figure}

Figure \ref{fig:tunnel_diode_omega_vs_eigenvalues_limit_cycle} shows the imaginary part of the complex frequency and the eigenvalues of the linearized system at each time step.  $|\omegau{u}|$ oscillates throughout the entire trajectory.  In this case, the elements of the Jacobian matrix of the circuit never converge to constant values and the eigenvalues lack of any physical or geometrical meaning.  As a matter of fact, in the period defined by the limit cycle, the eigenvalues switch from a pair of complex conjugate values to two real and then back again to complex conjugate.  The transformation that leads to calculate the eigenvalues of the system thus, apart from being time-dependent, is also non-isomorphic.  The geometrical meaning of $\rho$ and $|\tilde{\bs{\omega}}|$, on the other hand, is always that of rate of change of the radius and of curvature, respectively, of the closed curve described by the states.

\subsubsection{Two Coexisting Stable Equilibria}

In this scenario, we consider the case that the tunnel diode system has two stable equilibrium points. To achieve this, we use the following parameters:
\begin{itemize}
    \item Inductance: $L = 1.0$ H
    \item Capacitance: $C = 0.5$ F
    \item Resistance:  $R = 1.5 \ \Omega$
    \item Voltage Source: $V_{DC} = 0.35 $ V
\end{itemize}

The equilibrium point to which the system converges depends on the system's initial conditions. To showcase this, we examine and compare two cases with different initial conditions:
\begin{itemize}
    \item Case 1: $i_L(t=0) =  0$ A and $v_C(t=0)= 0$ V.  
    \item Case 2: $i_L(t=0) =  0$ A and $v_C(t=0) = 0.35$ V.  
\end{itemize}

\begin{figure}[htb]
  \centering
  \includegraphics[width=0.99\linewidth]{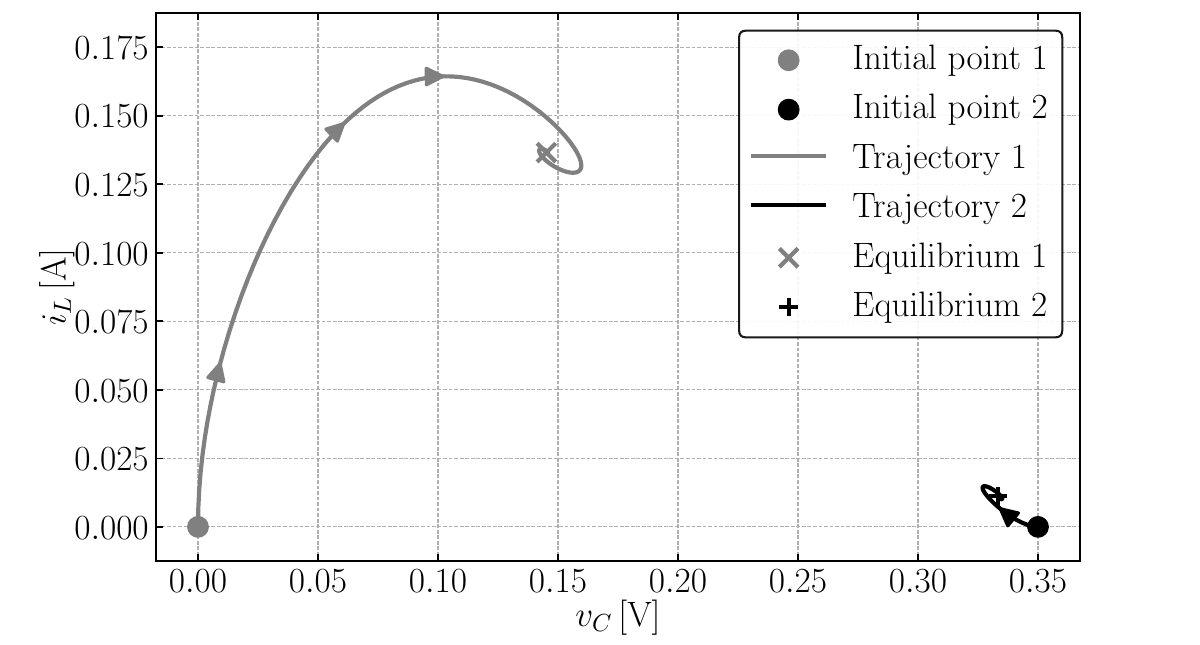}
  \caption{State space trajectories of the original states for a tunnel diode system with two equilibrium points.}
  \label{fig:tunnel_diode_phase_x_two_equilibria}
  \vspace{-2mm}
\end{figure}

\begin{figure}[htb]
  \centering
  \includegraphics[width=0.99\linewidth]{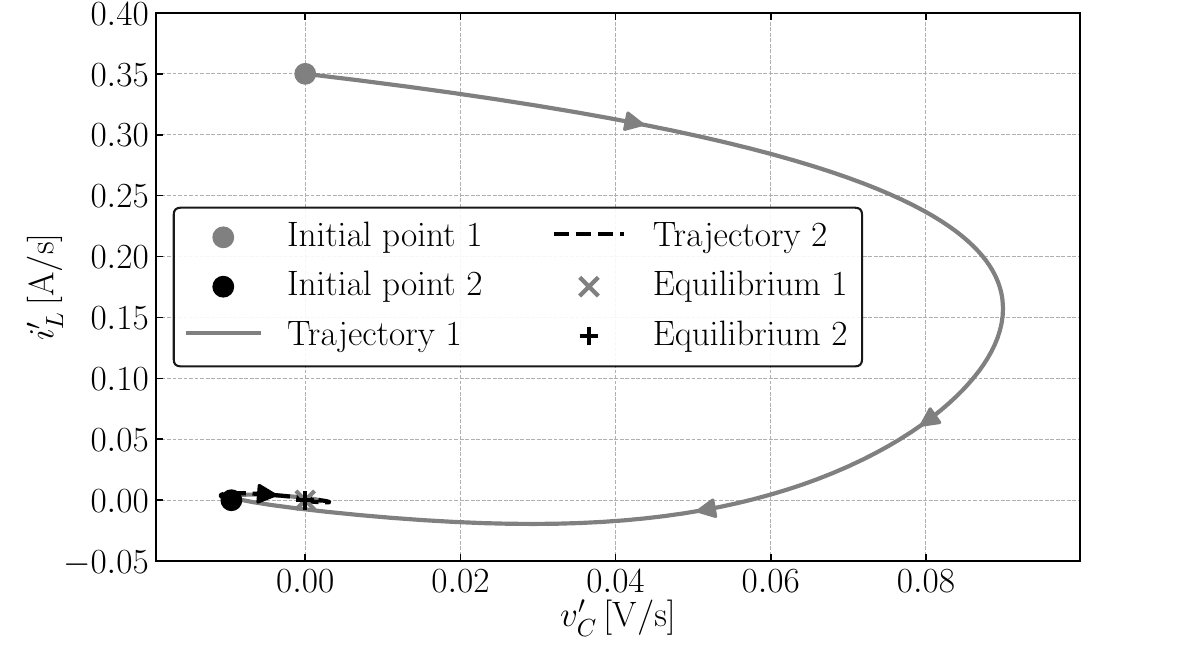}
  \caption{State space trajectories of the generalized velocities for a tunnel diode system with two equilibrium points.}
  \label{fig:tunnel_diode_phase_u_two_equilibria}
  \vspace{-2mm}
\end{figure}

Figure \ref{fig:tunnel_diode_phase_x_two_equilibria} shows the trajectories of the original states of the system for both cases, while Fig.~\ref{fig:tunnel_diode_phase_u_two_equilibria} shows the state space trajectories of the generalized velocities.  The two systems settle at different equilibrium points. However, since the time derivatives of all states become zero at equilibrium, both trajectories of generalized velocities converge to $(0,0)$.


\begin{figure}[htb]
  \centering
  \includegraphics[width=0.99\linewidth]{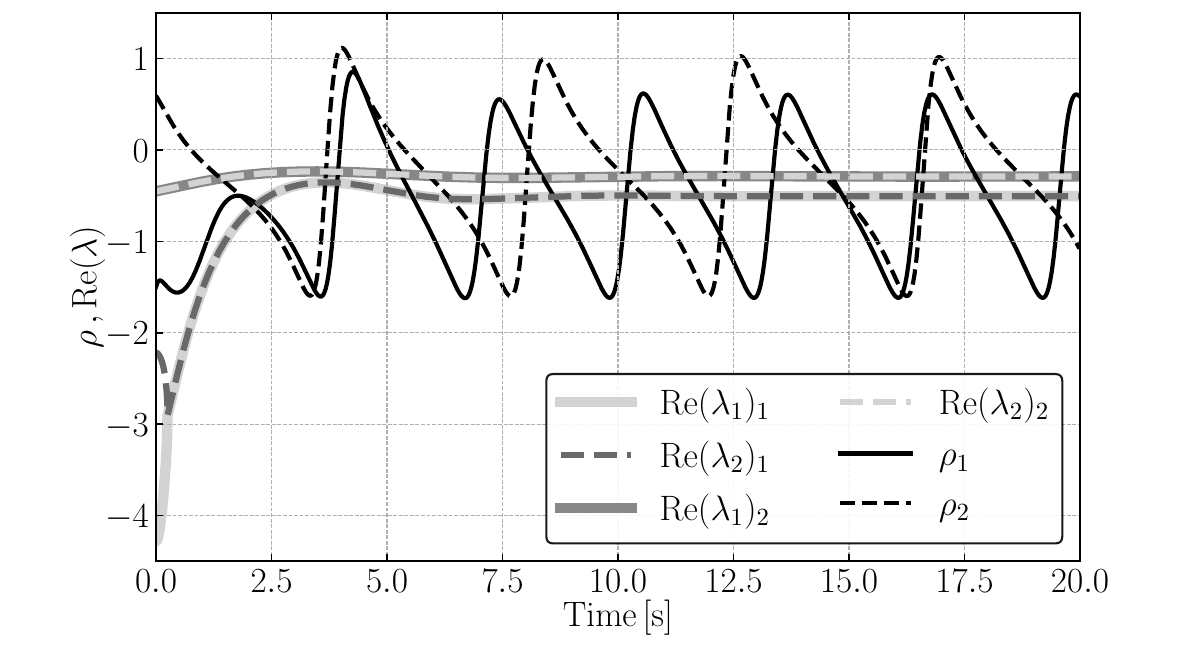}
  \caption{Imaginary part of complex frequency and of the eigenvalues of a tunnel diode system with two equilibrium points.}
  \label{fig:tunnel_diode_rho_vs_eigenvalues_two_equilibria}
  \vspace{-2mm}
\end{figure}

Figure \ref{fig:tunnel_diode_rho_vs_eigenvalues_two_equilibria} shows the real part of the complex frequency together with the real parts of the eigenvalues of the linearized systems at each time step. Since the systems in the two cases converge to different equilibrium points, the eigenvalues of their corresponding linearized systems are also different. In both cases, $\rho$ oscillates around the real part of the corresponding eigenvalue of each equilibrium point.

\begin{figure}[htb]
  \centering
  \includegraphics[width=0.99\linewidth]{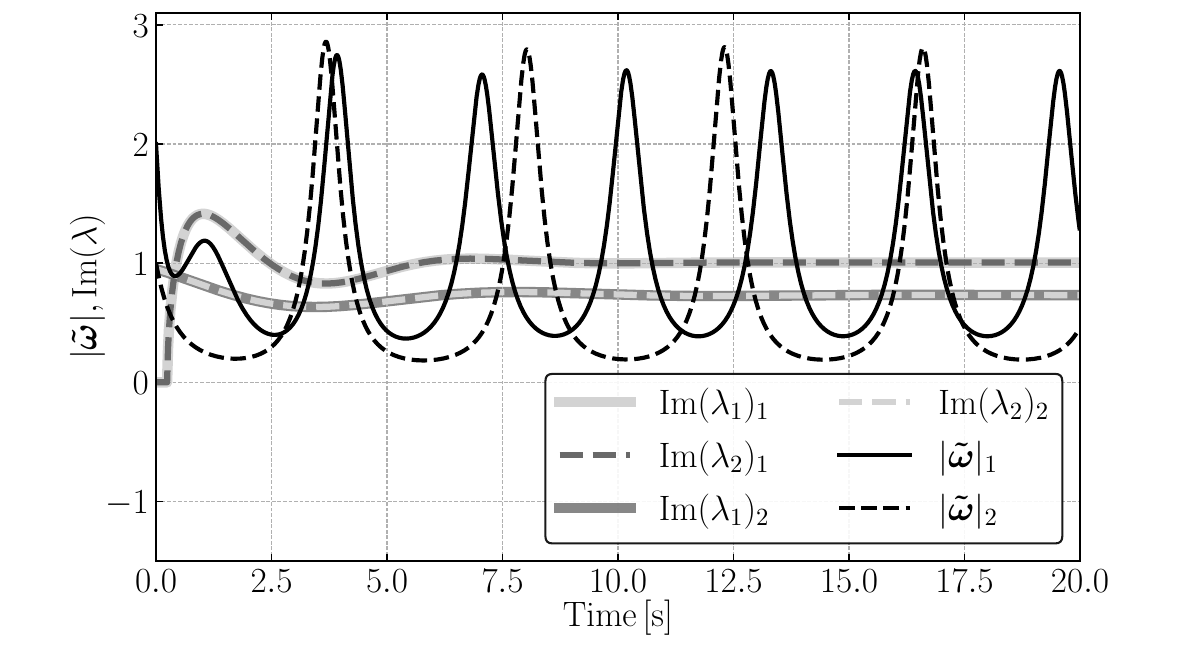}
  \caption{Imaginary part of complex frequency and of the eigenvalues of a tunnel diode system with two equilibrium points.}
  \label{fig:tunnel_diode_omega_vs_eigenvalues_two_equilibria}
  \vspace{-2mm}
\end{figure}

Figure \ref{fig:tunnel_diode_omega_vs_eigenvalues_two_equilibria} shows the imaginary part of the complex frequency together with the imaginary parts of the eigenvalues of the linearized systems at each time step. Since the systems in the two cases converge to different equilibrium points, the eigenvalues of their corresponding linearized systems are also different. In both cases, $|\omegau{}|$  oscillates around the imaginary part of the  corresponding eigenvalue of each equilibrium point. 

Figures \ref{fig:tunnel_diode_rho_vs_eigenvalues_two_equilibria} and \ref{fig:tunnel_diode_omega_vs_eigenvalues_two_equilibria} indicate that in both cases the systems have a pair of complex conjugate eigenvalues as they approach equilibrium, even though, at the beginning of its trajectory, the linearized system in Case 1 exhibits two real eigenvalues for a short period of time.

\section{Conclusions}
\label{sec:conclusions}

This work highlights that complex frequency and eigenvalues are equivalent descriptions of the same underlying dynamical properties when applied to \ac{lti} systems.  
The equivalence established in this paper is illustrated through a variety \ac{lti} systems and shows that eigenvalues are the complex frequencies of a non-isomorphic transformation that diagonalizes the state matrix of the system and decomposes it into a set of isotropic states.  Unifying algebraic stability criteria with the differential geometry of state-space trajectories provides relevant insights on the  behavior of the trajectories of physical quantities such as currents and voltages.

For nonlinear systems, this equivalence also holds if the system converges to an equilibrium.  In fact, for trajectories sufficiently close to an equilibrium, namely, for an almost constant Jacobian matrix, eigenvalues still represent a transformation of the states and converge to a steady state value.  However, if the system trajectory does not reaches an equilibrium and, for example, ends up on a limit cycle, eigenvalues lacks a geometrical and physical interpretation, whereas the invariants of the geometric frequency still retain their geometrical meaning, allowing for the possibility of extracting meaningful information about the system trajectory.

In summary, geometric frequency always captures instantaneous dynamical features that eigenvalues alone cannot represent.  Geometric frequency, thus, can be viewed as a mathematical object that naturally extends the concept of eigenvalues beyond an equilibrium-based analysis. 

Future work will focus on extending the proposed framework to nonlinear time-varying systems and investigating the properties of the bivector $\tilde{\bs{\omega}}$ for $n$-dimensional systems, especially power systems.  Furthermore, we intend to investigate the case in which the state matrix $\A$ is not diagonalizable, that is, it has at least one eigenvalue whose geometric multiplicity is strictly smaller than its algebraic multiplicity. Finally, we plan to study the possible relation between geometric frequency and the Lyapunov exponents of systems exhibiting chaotic motion.

\appendices

\section{Vector Operations}
\label{app:Operations}

In this appendix, the vector operations that were used throughout the paper for the calculation of geometric and complex frequency are presented. 

Let $\x = (x_1, x_2, ..., x_n)$ and $\y = (y_1, y_2, ..., y_n)$  be two $n$-dimensional vectors in $\mathbb{R}^n$.

The \textit{inner product} is defined as:
\begin{equation}
    \x \cdot \y = \sum_{i=1}^n x_i y_i
    \label{eq:inner_product}
\end{equation}
In this work, the highest dimension in which the inner product is in $\mathbb{R}^2$, where for example $\x \cdot \y = x_1y_1 + x_2y_2$. The inner product is symmetric, associative and communicative. The magnitude of $\x$ is defined as:
\begin{equation}
    |\x| = \sqrt{\x \cdot \x} \,.
    \label{eq:magnitude}
\end{equation}

The \textit{outer product} is defined as:
\begin{equation}
    \x \otimes \y = \begin{bmatrix}
    x_1 y_1 & \cdots & x_1 y_n \\ 
    \vdots  & \ddots & \vdots  \\ 
    x_n y_1 & \cdots & x_n y_n
    \end{bmatrix} \,.
\end{equation}

The \textit{wedge product} is defined as:
\begin{equation}
    \x \wedge \y = \x \otimes \y - \y \otimes \x \,.
\end{equation}

Specifically, in $\mathbb{R}^2$, which is the dimension that  the wedge product is used throughout this paper, $\x \wedge \y = (x_1y_2 - x_2y_1) \bs{e}_1\wedge\bs{e}_2$, where $\bs{e}_1 = \begin{bmatrix}
    1 \\ 0
\end{bmatrix}$ and $\bs{e}_2 = \begin{bmatrix}
    0 \\ 1
\end{bmatrix}$ are the unit vectors that consist the basis of $\mathbb{R}^2$.

The \textit{geometric product} is defined as:
\begin{equation}
    \bs{xy} = \x \cdot \y + \x \wedge \y \,.
\end{equation}
The result of the geometric product, which is called \textit{multivector}, consists of two components. The first component, $\x \cdot \y$, is a scalar that represents the projection of $\y$ onto the vector $\x$. The second component, $\x \wedge \y$, represents a bivector orthogonal to the space defined by the vectors $\x$ and $\y$. 

\section{Derivation of $\bar{\bs{\xi}}' = \bar{\bs{\Lambda}} \bar{\bs{\xi}}$}
\label{app:xi_dynamics}

In this appendix, the proof that \eqref{eq:xi} leads to the dynamic system in \eqref{eq:xi_dynamic} is presented. 

Starting with \eqref{eq:xi} and
differentiating both sides with respect to time gives:
\begin{equation}
\label{eq:derivation_1}
    \bs{\bar{\xi}}' = \bar{\bs{U}}' \bs{u} + \bar{\bs{U}} \bs{u}'\, .
\end{equation}
Since the system is time-invariant, the matrix $\bar{\bs{U}}$ is constant, so its time derivative is: $\bar{\bs{U}}' = \mathbf{0}$. Therefore:
\begin{equation}
\label{eq:derivation_2}
    \bs{\bar{\xi}}' = \bar{\bs{U}} \bs{u}'\, .
\end{equation}
Substituting the original linear dynamics defined in \eqref{eq:system_speed} yields:
\begin{equation}
\label{eq:derivation_3}
    \bs{\bar{\xi}}' = \bar{\bs{U}} \bar{\bs{A}} \bs{u}\, .
\end{equation}
Then, solving \eqref{eq:xi} for $\bs{u}$ gives: 
\begin{equation}
\label{eq:solve_for_u}
    \bs{u} = \bar{\bs{U}}^{-1} \bs{\bar{\xi}} \, .
\end{equation}
Replacing \eqref{eq:solve_for_u} in \eqref{eq:derivation_3} yields:
\begin{equation}
\label{eq:derivation_4}
    \bs{\bar{\xi}}' = \bar{\bs{U}} \bar{\bs{A}} \left( \bar{\bs{U}}^{-1} \bs{\bar{\xi}} \right) 
    = \left( \bar{\bs{U}} \bar{\bs{A}} \bar{\bs{U}}^{-1} \right) \bs{\bar{\xi}}\, .
\end{equation}
By construction, from \eqref{eq:decomposition}, $\bar{\bs{U}}$ diagonalizes $\A$:
\begin{equation}
    \bar{\bs{U}} \bar{\bs{A}} \bar{\bs{U}}^{-1} = \bar{\bs{\Lambda}}\, .
\end{equation}
Substituting this relation to \eqref{eq:derivation_4}, we obtain :
\begin{equation*}
    \bs{\bar{\xi}}' = \bar{\bs{\Lambda}} \bs{\bar{\xi}} \, .
\end{equation*}

We have thus shown, how we can derive the dynamic system of \eqref{eq:xi_dynamic} from \eqref{eq:xi}.



\vfill

\end{document}